\documentclass[prb,twocolumn,superscriptaddress,showpacs,amsmath,amssymb,longbibliography]{revtex4-2}

\usepackage{graphicx}% Include figure files
\usepackage{dcolumn}% Align table columns on decimal pointb
\usepackage{bm}% bold math
\usepackage{gensymb}
\usepackage{multirow}
\usepackage[none]{hyphenat}
\usepackage{epstopdf}
\usepackage{float}
\usepackage{subfigure}
\usepackage{tabularx}
\usepackage{color}
\usepackage{siunitx}
\usepackage{xr}
\usepackage{amsmath} 
\usepackage{mathtools}
\usepackage{braket}
\usepackage[normalem]{ulem}
\usepackage{natbib}
\usepackage[caption=false]{subfig}
\usepackage{afterpage}
\usepackage{enumerate}
\usepackage{blindtext}
\makeatletter
\newcommand*{\rom}[1]{\expandafter\@slowromancap\romannumeral #1@}
\makeatother

\begin{document}

\newcommand{\lsy}[1]{{\color{red} #1}}

\newcommand{\sr}[1]{{\color{magenta} #1}}
\newcommand{\beq}{\begin{equation}}
\newcommand{\eeq}{\end{equation}}
\newcommand{\vS}{\vec{S}}
\newcommand{\ud}{\mathrm{d}}

%\title{Investigations of \emph{order by projection} in the Single-band Hubbard Model Using Density Matrix Renormalization Group}
\title{Order by projection in single-band Hubbard model: a DMRG study}
\author{Shuyi Li}
\affiliation{Department of Physics, University of Florida, Gainesville, FL 32611, USA}

\author{Cheng Peng}
\affiliation{Stanford Institute for Materials and Energy Sciences, Menlo Park, CA 94025, USA}

\author{Yue Yu}
\affiliation{Department of Computer Science, University of Florida, Gainesville, FL 32611, USA}

\author{B. Sriram Shastry}
\affiliation{Department of Physics, University of California, Santa Cruz, CA 95064, USA}

\author{Chunjing Jia}
\email[Correspondence e-mail address: ]{chunjing@phys.ufl.edu}
\affiliation{Department of Physics, University of Florida, Gainesville, FL 32611, USA}

\date{\today}

\begin{abstract}

In a Fermi system near or at half-filling, a specific superconducting pairing channel, if not explicitly included in the Hamiltonian, can be boosted by suppressing a competing pairing channel; this is exemplified by the enhancement of extended $s$-wave correlations upon suppressing $s$-wave Cooper pairing. This phenomenon, originally found by the use of generalized uncertainty relations is referred to as \emph{order by projection}. The case of zero on-site Coulomb interaction in the thermodynamic limit, confirms this mechanism through the analytical solution. In this study, we go further and systematically investigate this mechanism for a strongly correlated fermionic Hubbard model, now with finite on-site interaction, on a square lattice with an extended set of hopping parameters. We explore the behaviors of different pairing channels when one of them is suppressed, utilizing density matrix renormalization group calculations. Our findings provide numerical evidence supporting the existence of \emph{order by projection} in the strongly correlated system we studied. We also investigate the effect of the strength of Hubbard $U$, next-nearest neighbor $t'$, hole-doping, as well as finite-size scaling approaching the thermodynamic limit.
\end{abstract}

%\pacs{xxx}

\maketitle

\vspace{8mm}
\noindent
\MakeUppercase{\textbf{Introduction}}

The Hubbard model~\cite{Hubbard} stands as one of the foundational pillars in the research of strongly correlated electron systems, illuminating phenomena ranging from Mott-insulator to magnetism. More importantly, the Hubbard model and its extensions have been one of the frequently used low-energy effective models for interpreting the high-temperature superconductivity in transition-metal oxides~\cite{Dagotto}. Unlike conventional superconductivity, which are well-described by Bardeen–Cooper–Schrieffer (BCS) theory and typically involve electron pairs bound through lattice vibrations~\cite{BCS1,BCS2}, unconventional superconductivity arises from mechanisms beyond the BCS framework, such as electronic correlation and spin fluctuations~\cite{Keimer, LeTacon}. However, there are widespread debates about whether the Hubbard model encompasses a superconducting phase and can be consequently applied to unconventional superconductivity, depending on individual beliefs and the methods used. Addressing the gaps in the model study and bridging the discrepancy between the understanding of the Hubbard model and its ability to faithfully explain all experimental phenomena is challenging. This discrepancy is constrained by the model's inherent inability to be solved exactly, both theoretically and numerically. As a result, reaching a conclusion that achieves the broadest possible consensus remains difficult and is still under development. %As a limiting case, the feasibility of purely electronic repulsive interaction-driven superconductivity continues to be an important open question and is of considerable interest for its possibility to explain the high-$T_c$ superconductors. {\color{blue} (the word ``feasibility" makes me think it may not be feasible.)}

In our work, we explore one feasible method for tuning superconductivity in the single-band Hubabrd model~\cite{orderbyprojection1}, which involves the application of the uncertainty principle~\cite{uncertaintyP}. %which is a fundamental concept in quantum mechanics. ({\color{blue} CJ: ``Multiple types of electronic interactions" sound disconnected to the later part of the sentence.}) 
This principle provides a lower bound limitation on the product of the fluctuations of two conjugate physical quantities, the most famous example was expressed succinctly by Heisenberg's inequality $\Delta x\Delta p \geq \hbar/2$. In strongly correlated electron systems, it has been found that the Cooper pairs with different symmetries could form pairs of ``conjugate quantities", such as the extended $s$-wave Cooper pairs and the $s$-wave Cooper pairs on a $d$-dimensional hypercubic lattice~\cite{orderbyprojection1}. In analogy to the inequality of position and momentum, the pairing correlations of extended $s$-wave Cooper pairs could be enhanced from the process of suppressing the $s$-wave Cooper pairs, which forms a ``superconductivity squeezed state". This is what we called \emph{order by projection}, a phenomenon originally introduced by Shastry~\cite{orderbyprojection1,orderbyprojection2,orderbyprojection3,orderbyprojectionexact}.

On the framework of the single-band Hubbard model, \emph{order by projection} is realized in its simplest form by adding a term to the Hamiltonian (as will discussed in Eq.~\ref{Eq-1}) that amounts to projecting out $s$-wave Cooper pairs. %As noted above, in the case of the Hubbard model with $U=0$, an exact analytical solution is available at half-filling and low doping in the thermodynamic limit~\cite{orderbyprojectionexact}. A notable feature is its exhibition of quasi-long-range order at half-filling, where a certain correlation function grows as the $1.5$ power of the system size, rather than linearly. However, 
When \emph{order by projection} inequalities are applied to the Hubbard model with finite on-site repulsion $U$, as in the case of cuprates where the Hubbard $U$ can be relatively strong, only a lower bound of the extended $s$-wave Cooper pairs can be provided analytically. Whether the \emph{order by projection} mechanism can be broadly applied to finite $U$ systems with different superconducting pairing symmetries is beyond the scope of the analytical description and has not been studied yet. Targeting the above gaps, we use a numerical unbiased method, density matrix renormalization group (DMRG)~\cite{White1992}, to study an extension of the Hubbard model that includes a projection term of the $s$-wave Cooper pairs.

\vspace{8mm}
\noindent
\MakeUppercase{\textbf{Results}}
\\
\textbf{Model}

\begin{figure}[ht!]
\centering
\includegraphics[width=\linewidth]{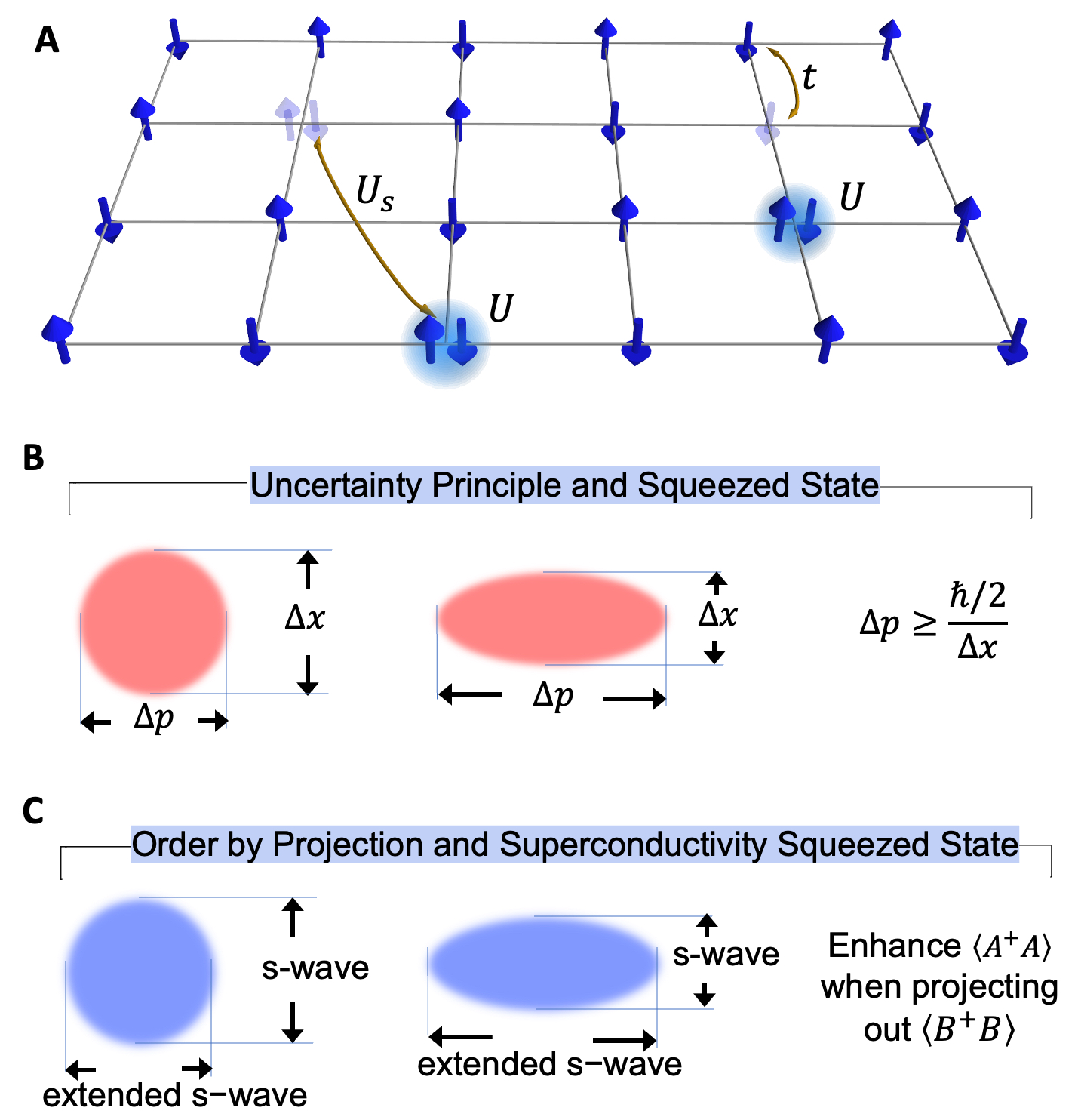}
\caption{Model Hamiltonian and analogue of order by projection to the uncertainty principle. (a) A schematic of the single-band Hubbard model with the projection out of the $s$-wave pairing for a square lattice in our calculation. $U$ represents the on-site Coulomb interaction for double occupancy. One gains energy $t$ for single electron hopping between nearest neighbors; one pays energy $U_s$ for pair-hopping, where the pair-hoppings pertain to long range. The next nearest neighbor hopping $t'$ is not shown in this schematic. (b) Uncertainty principle and squeezed state for space and momentum. (c) Order by projection mechanism and ``superconductivity squeezed state" for $s$-wave pairing ($\hat{B}$) and extended $s$-wave pairing ($\hat{A}$). The extended $s$-wave pairing is enhanced when projecting out the $s$-wave pairing.}
\label{fig:modelHamiltonian}
\end{figure}

We are interested in the many-body fermionic system on a square lattice, as shown in Fig.~\ref{fig:modelHamiltonian}, with the projection out of the $s$-wave Cooper pairs, described by the Hamiltonian as:
\begin{equation}\label{Eq-1}
    \hat{H}=\hat{H}_{0}+\hat{H}_{\text{proj}}~. 
\end{equation}
The first term $\hat{H}_{0}$ is the single-band fermionic Hubbard model, defined as:
\begin{equation}
\begin{aligned}
   \hat{H}_{0}=&-t\sum_{\langle ij\rangle,\sigma}(\hat{c}^{\dagger}_{i\sigma}\hat{c}_{j\sigma}+\text{h.c.})\\&-t'\sum_{\langle\langle ij\rangle\rangle,\sigma}(\hat{c}^{\dagger}_{i\sigma}\hat{c}_{j\sigma}+\text{h.c.})+U\sum_{i}\hat{n}_{i\uparrow}\hat{n}_{i\downarrow},
\end{aligned} \label{Eq-2}
\end{equation}
where $\hat{c}_{i\sigma}^{\dagger}$ $(\hat{c}_{i\sigma})$ is the creation (annihilation) operator of an electron with spin $\sigma=\uparrow,\downarrow$ at site $i$, $t$ $(>0)$ %{\color{red} (CP: you could define $t$ as an energy unit and use $U=...$ instead of $U=...t$.)} 
and $t'$ are hopping integrals between the nearest neighbor and the next-nearest neighbor sites, $U$ is the on-site Coulomb repulsion, and $\hat{n}_{i\sigma}=\hat{c}^{\dagger}_{i\sigma}\hat{c}_{i\sigma}$ is the number operator. The Hubbard model with on-site repulsion $U/t\sim8$ is believed to describe the low energy physics relevant to %high-temperature superconducting 
cuprates. For simplicity, we set $t$ as the energy unit in the following discussions.

The second term $\hat{H}_{\text{proj}}$ penalizes the hoppings of the $s$-wave Cooper pairs with energy $U_s$:
\begin{equation}
    \hat{H}_{\text{proj}}=U_s \hat{B}^{\dagger}\hat{B}, \label{Eq-3}
\end{equation}
where $\hat{B}$ is the annihilation operator for the $s$-wave Cooper pairs summed over all sites: %({\color{blue} Are $\hat{B}$ and $B$ different?} {\color{red} CP: I added $\hat{}$ on all operators.})

\begin{equation}
    \hat{B}=\sum_{i}\hat{c}_{i\downarrow}\hat{c}_{i\uparrow}~.
\end{equation}

If $U_s$ is negative, it encourages the formation of $s$-wave pairing in the ground state. This is simply understood because at the mean field level, Eq.~(\ref{Eq-3}) reduces to the BCS pairing Hamiltonian. Conversely, and less intuitively, it has been shown that a positive $U_s$ which strongly discourages $s$-wave pairing, leads instead to the enhancement of the extended $s$-wave pairing. This is the so-called \emph{order by projection} effect.\\

%This phenomenon is called \emph{order by projection} and was first introduced by Shastry~\cite{orderbyprojection1}. In the case of Hubbard model with Coulomb repulsion $U=0$, an exact solution of the enhancement of extended $s$-wave pairing and the suppression of $s$-wave pairing at half-filling and low doping in the thermodynamic limit has been provided analytically~\cite{orderbyprojectionexact}. Particularly, it exhibits quasi-long-range order at half-filling. 

%However, whether the ``order by projection" scenario still applies to finite $U$ (around $8t$) Hubbard model, and to what extent it still applies, are beyond the scope of analytical description, thus have not been studied yet. %In this work, we investigate this problem numerically using density matrix renormalization group (DMRG) method, which is a numerically exact method to solve ground state of strong electronic system in relatively large system size. 
%\noindent
%\textbf{Pair-Density Matrices}
%\\
%\subsection{Pair-Density Matrices}
The main objective of this work is to study the influence of the projection term $\hat{H}_{\text{proj}}$ on different channels of Cooper pairs. Here we list the Cooper pairs considered in this work: extended $s$-wave pairing ($\hat{A}$), $s$-wave pairing ($\hat{B}$), and $d$-wave pairing ($\hat{D}$) on the square lattice, which are defined as follows:
\begin{equation}
\begin{aligned}
    \hat{A}&=\sum_i \hat{A}_i=\sum_i \hat{\Delta}_{i,i+\hat{x}}+\hat{\Delta}_{i,i+\hat{y}},\\
    \hat{B}&=\sum_i \hat{B}_i=\sum_i \hat{c}_{i\downarrow}\hat{c}_{i\uparrow},\\
    \hat{D}&=\sum_i \hat{D}_i=\sum_i \hat{\Delta}_{i,i+\hat{x}}-\hat{\Delta}_{i,i+\hat{y}},\\
\end{aligned}
\end{equation}
where $\hat{\Delta}_{i,j}=\hat{c}_{i\uparrow}\hat{c}_{j\downarrow}+\hat{c}_{j\uparrow}\hat{c}_{i\downarrow}$. The correlation functions related to these pairings are called pair-density matrices:
\begin{equation}\label{eq:pairdensitymatrices}
P_{ij}^A=\langle \hat{A}_i^{\dagger}\hat{A}_j \rangle,~P_{ij}^B=\langle \hat{B}_i^{\dagger}\hat{B}_j \rangle,~P_{ij}^D=\langle \hat{D}_i^{\dagger}\hat{D}_j \rangle,
\end{equation}
where $\langle \cdots\rangle$ denotes the expected value at the ground state $|\Phi_0\rangle$ of the Hamiltonian described in Eq.(\ref{Eq-1}).

We first evaluate the average Cooper pair densities across all sites ($N_s$):
\begin{equation}
\begin{aligned}
    \rho_A=\frac{\langle \hat{A}^{\dagger}\hat{A}\rangle}{N_s}&= \frac{1}{N_s}\sum_{i,j}P_{ij}^A,\\
    \rho_B=\frac{\langle \hat{B}^{\dagger}\hat{B}\rangle}{N_s}&= \frac{1}{N_s}\sum_{i,j}P_{ij}^B,\\
    \rho_D=\frac{\langle \hat{D}^{\dagger}\hat{D}\rangle}{N_s}&= \frac{1}{N_s}\sum_{i,j}P_{ij}^D
\end{aligned}
\end{equation}
to measure the enhancement or suppression of the Cooper pairs with different pairing symmetries.  Additionally, we will explore the eigenvalues of the pair-density matrices in Eq.~(\ref{eq:pairdensitymatrices}), in which the largest eigenvalue of each pair-density matrix corresponds to the condensate occupation of each order~\cite{PenroseOnsager}. If a specific kind of Cooper pair condenses, its condensation occupation scales linearly with the system size $\sim O(N_s)$ for a fixed electron density, while all other eigenvalues scale as $\sim O(1)$. This behavior indicates that the system exhibits off-diagonal long-range order~\cite{PenroseOnsager,CNYang}. Thus, we look at the ratio between the largest and second-largest eigenvalues of its pair-density matrix, which are denoted respectively for the extended $s$-wave, $s$-wave, and $d$-wave pairings as: 
\begin{equation}\label{Eq-8}
    R_A=\frac{\Lambda_1(P_{ij}^A)}{\Lambda_2(P_{ij}^A)},~R_B=\frac{\Lambda_1(P_{ij}^B)}{\Lambda_2(P_{ij}^B)},~R_D=\frac{\Lambda_1(P_{ij}^D)}{\Lambda_2(P_{ij}^D)}.
\end{equation} 
The ratio $R$ was initially introduced by Rigol, Shastry, and Haas~\cite{Shastry_Ratio1,Shastry_Ratio2}. Calculating the ratio $R$ is equivalent to studying the largest eigenvalue without accounting for any normalization effects. When condensation takes place, the behavior of the ratio is anticipated to follow $R\sim\Psi^2N_s+\Phi$ for a large number of sites $N_s$, where $\Psi$ represents the order parameter with an order of $O(1)$, and $\Phi$ exhibits sublinear scaling with respect to the system size.

We aim to investigate that whether turning on pair hoppings $U_s$ atop the Hubbard model with finite $U$ will enhance or decrease pairings of interest, along with other relevant quantities to understand the underlying physics of the new states. We will investigate the ground state energy, ground state wavefunction and the corresponding pair density matrices for the model Hamiltonian using DMRG, which is a powerful numerical method to provide the ground state wavefunction with high resolution for strongly correlated systems at relatively large quasi-one-dimensional clusters. In our calculations, we use cylindrical clusters with the length and width of the system taken as $L_x$ and $L_y$, given the total site number $N_s=L_x\cdot L_y$. We use open boundary conditions along the $x$-axis and set $L_x$ to $4$, $8$, $16$, $20$, and $32$ for fine length scaling, and maintain periodic boundary conditions along the $y$-axis with $L_y$ fixed at $4$, which helps retain some properties beyond one dimension while reducing computational difficulty.\\

\noindent
\textbf{\emph{Order by projection} at half-filling}
\\

%{\color{red} (CP: 1. what's the purpose of showing the ground state energy per site $E/N_s$ seems not explicitly discussed. 2. I am confused about the ordering of the figures and sub figures, so I made a lot changes to .)}

The first question we want to address is whether the enhancement of the extended $s$-wave pairing, achieved by projecting out the $s$-wave pairing scenario, remains valid when the Hubbard repulsion is finite at half-filling. To gain an overall understanding of this question, we first explore the impact of the repulsive interaction, within the range of $0 < U_s \leq 1$, which projects out the $s$-wave pairing. 

\begin{figure}[htbp!]
\centering
\includegraphics[width=0.9\linewidth]{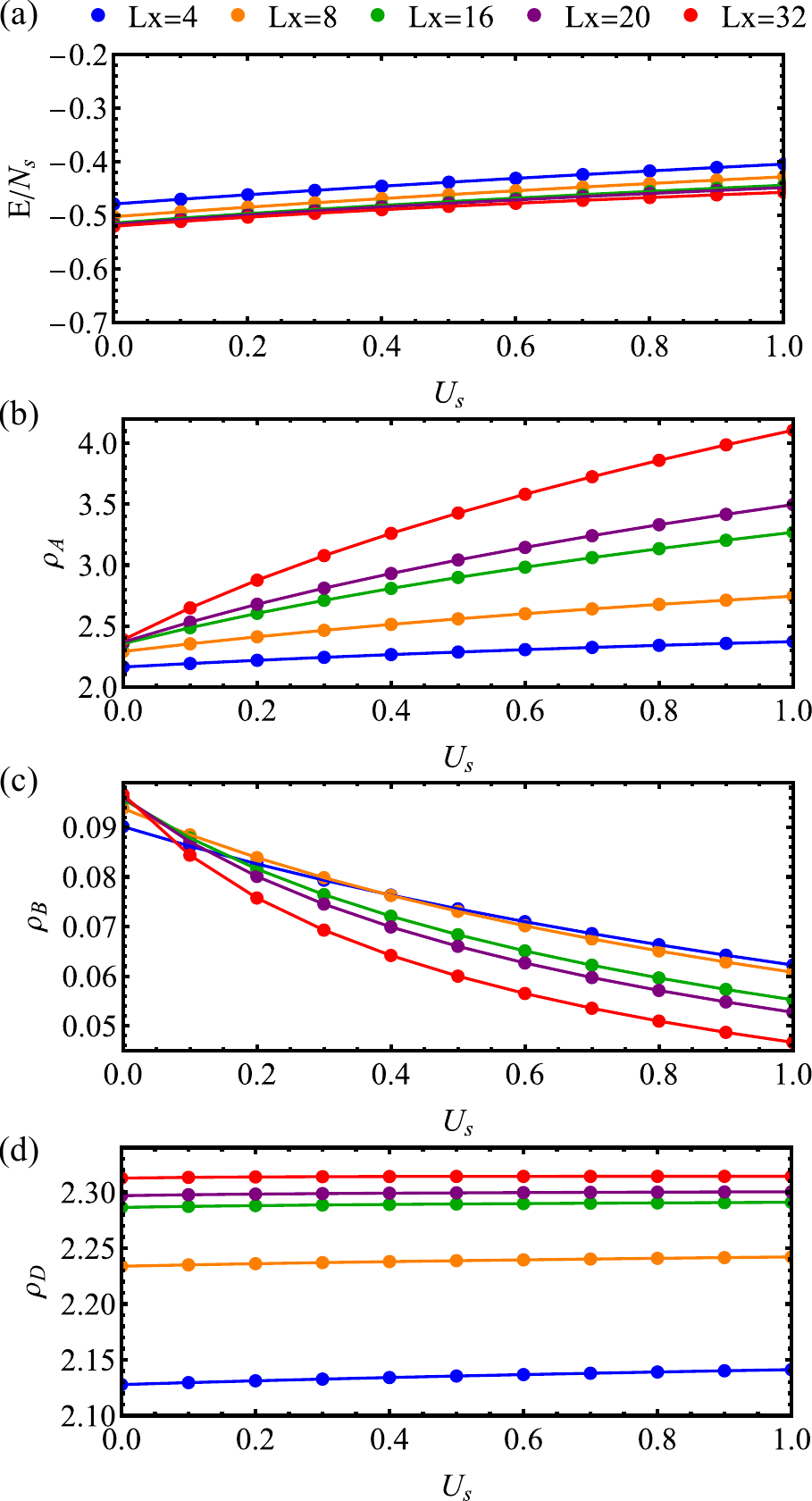}
\caption{(a) Ground state energy $E/N_s$, (b) extended $s$-wave pair density $\rho_A$, (c) $s$-wave pair density $\rho_B$ and (d) $d$-wave pair density $\rho_D$ versus strength $U_s$ of projection term $\hat{H}_{\text{proj}}$ under $U=8$ and $t'=0$ at half-filling. System length $L_x=4,8,16,20,32$ are tested.}
\label{fig:result(U8t'0)}
\end{figure}

Fig.~\ref{fig:result(U8t'0)} shows the $U_s$ dependence of ground state energy per site $E/N_s$, pair density of extended $s$-wave $\rho_A$, pair density of $s$-wave $\rho_B$, and pair density of $d$-wave $\rho_D$ %(in units of $t$)
for multiple system sizes at half-filling. Results for weak attraction with a negative $ U_s $ within the range of $-1/N_s \leq U_s < 0$ are provided in the supplementary material~\cite{splm} for completeness.\\

\begin{figure}[htbp!]
\centering
\includegraphics[width=0.9\linewidth]{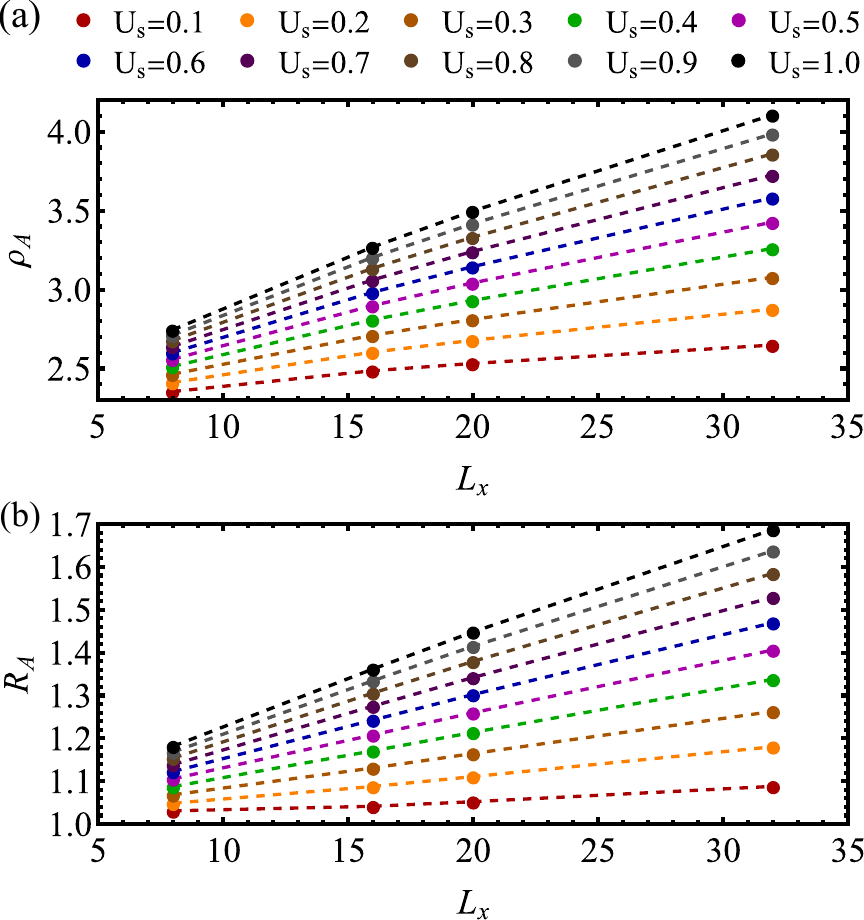}
\caption{(a) Extended $s$-wave pair density $\rho_A$ and (b) the ratio $R_A$ between the largest and second-largest eigenvalues of the extended $s$-wave pair-density matrix (see Eq.~\ref{Eq-8}) versus system length $L_x$ in the case of $U=8$ and $t'=0$ at half-filling for $U_s=0.1\sim 1$.}
\label{fig:result(U8t'0)R}
\end{figure}

Fig.~\ref{fig:result(U8t'0)} (a) shows the change in ground state energy per site is as small as $\sim 0.04$ when $U_s$ varies from $0$ to $1$ and diminishes as $L_x$ increases. This scale of energy change can be considered a perturbation to $H_0$. The enhancement of the extended $s$-wave pair density $\rho_A$ in Fig.~\ref{fig:result(U8t'0)} (b) confirms the prediction of the \emph{order by projection} mechanism. In the same $U_s$ region, the $s$-wave pair density $\rho_B$ is suppressed with increasing $U_s$, as shown in Fig.~\ref{fig:result(U8t'0)} (c), reflecting the ``projection out'' of the $s$-wave pairing. %In the case of $L_x = 32$, $\rho_A$ increases by approximately $70\%$, and $\rho_B$ decreases by around $50\%$, when $U_s$ varies from $0$ to $t$. 
Additionally, the impact of $U_s$ on the enhancement of $\rho_A$ and the suppression of $\rho_B$ becomes more pronounced as the system size $N_s$ increases. The dependence of $\rho_A$ on the cylinder length $L_x$ is shown in Fig. \ref{fig:result(U8t'0)R} (a). Assuming that the leading order of $\rho_A$ is proportional to $L_x^{\alpha}$, we fitted the calculated data for $0.1 \leq U_s \leq 1$ and obtained an index $\alpha$ in the range of $0.3 < \alpha < 0.6$. The scaling of $\rho_A$ is equivalent to $\langle A^{\dagger}A\rangle \propto L_x^{1+\alpha}$, which indicates that the system exhibits quasi-long-range order. The $d$-wave pair density $\rho_D$, however, remains almost unchanged, as shown in Fig.~\ref{fig:result(U8t'0)}(d), indicating that the $d$-wave pair density is not sensitive to \emph{order by projection} of $s$-wave pairing and functions as an independent channel.\\

% Fig.2

Next, we calculate the ratios $R_A$, $R_B$, and $R_D$ defined in Eq.~\ref{Eq-8} as functions of $U_s$. Only $R_A$ increases monotonically with $U_s$, while $R_B$ and $R_D$ remain approximately $1$ and are almost unchanged for large system sizes (with the results of $R_A$ shown in Fig. \ref{fig:result(U8t'0)R}(b), and $R_B$ and $R_D$ shown in the supplementary material~\cite{splm}). If we consider the leading order of $R_A$ to be $L_x^{\beta}$, the fitted data for $0.1\leq U_s\leq 1$ in Fig.~3(b) show that the index $\beta$ stays in the range of $1.2>\beta>0.85$. This indicates that the extended $s$-wave Cooper pairs condense due to the \emph{order by projection} mechanism. 

%Additionally, we investigate the dependence of $R_A$ on the cylinder length $L_x$, as presented in Fig. \ref{fig:result(U8t'0)R}(b). \lsy{Change?} If we consider the leading order of $R_A$ to be $L_x^{\alpha}$, then $\alpha \gtrsim 1$ when $U_s < 0.3$ and $0 < \alpha \lesssim 1$ when $U_s > 0.3$. This indicates that the extended $s$-wave Cooper pairs condense due to the \emph{order by projection} mechanism.\\

% Fig.3

%Subsequently, we investigate whether the mechanism of {\it order by projection} will be influenced by the next-nearest neighbor hopping $t'$, as well as variations in Hubbard repulsion $U$ and the doping-dependence in the Hubbard model. Additionally, we examine finite size scaling in the strongly correlated regime $U=8t$ and analyze how the enhanced pairing scales with the system size.\\
%\subsection{$\mathbf{U=8t}$ at half-filling}

%\lsy{Furthermore, the effect of $U_s$ on the enhancement of $\rho_A$ and suppression of $\rho_B$, which is reflected by their slopes $|\partial \rho_A/\partial U_s|$ and $|\partial \rho_B/\partial U_s|$, becomes more remarkable as the system size $N_s$ increases, which is about $72\%$ $(-52\%)$ from $U_s=0$ to $U_s=t$ in the case of $L_x=32$, while the ground state energy increases less $(12\%)$ and the $d$-wave pair density $\rho_D$ changes very slightly. If $U_s$ is negative, $\rho_B$ $(\rho_A)$ is enhanced (suppressed) with $|U_s|$ increases, which is opposite to the previous case.} 

%{\color{red} CJ: big O means on the same order in computer science. It is a little bit confusing using big O symbol here. What are the values of $\alpha$ here? I would be nice to show the exact value of $\alpha$. What is the scaling for $U=0$ in Shastry's study? What is the significance of $R_A$ scaling almost linear of $L_x$? Why for the finite-size scaling, all the calcualtion are based upon small $U$. This seems to be disconnected with Figure 3(a) where $U$ is focused on $8t$.}    

\begin{figure}[htbp!]
\hspace{-1cm}
\centering
\includegraphics[width=0.9
\linewidth]{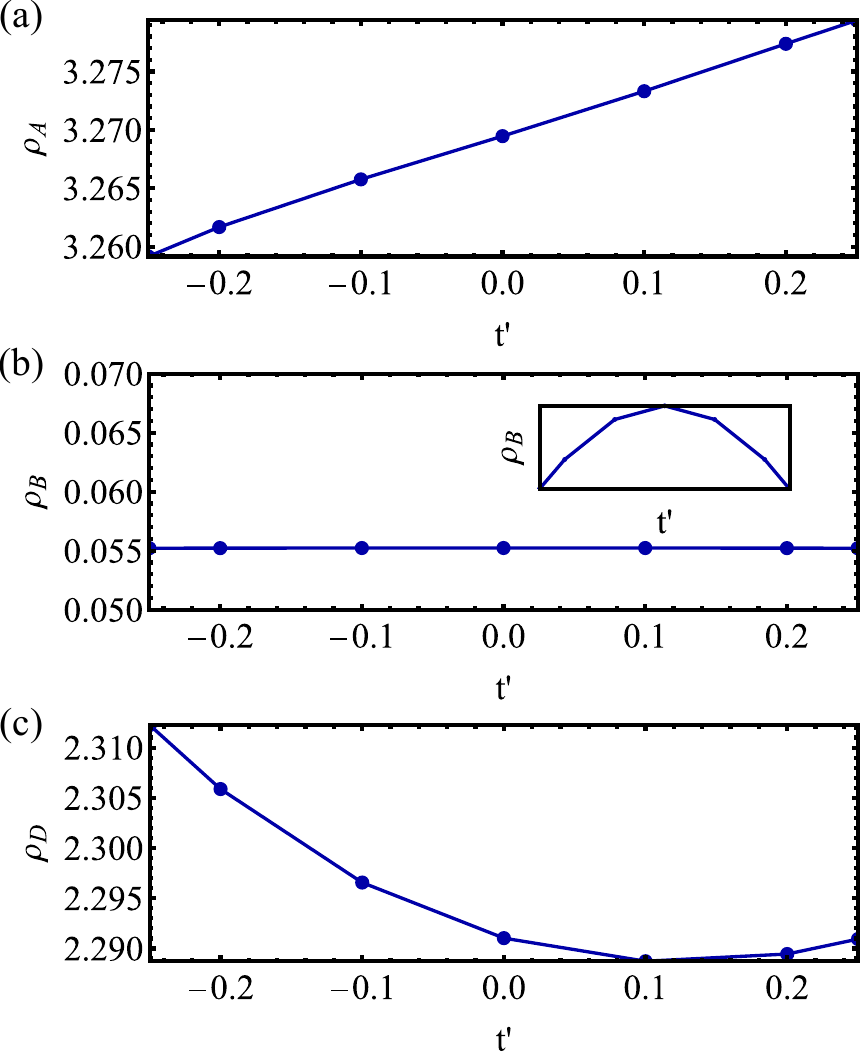}
\caption{(a) Extended $s$-wave pair density $\rho_A$, (b) $s$-wave pair density $\rho_B$ (the inset zooms in on the energy range by a factor of $10^3$) and (c) $d$-wave pair density $\rho_D$ versus the second nearest neighbor hopping $t'$ under $U=8$ and $U_s=1$ at half-filling.}
\label{fig:result(U8t')}
\end{figure}

Furthermore, we investigate whether the mechanism of {\it order by projection} will be influenced by the next-nearest neighbor hopping $t'$. Three kinds of pair densities calculated as functions of $t'$ under $U=8$ and $U_s=1$ are shown in Fig. \ref{fig:result(U8t')}. Generally speaking, the impact of $t'$ is marginal, as indicated by the limited range of values observed for these pair densities, yet we can still discern the different effect of $t'$ on the three kinds of pair densities. When $t'>0~(<0)$, $\rho_A$ increases (decreases) with continuously increasing $|t'|$, while $\rho_B$ remains scarcely affected by $t'$ for the same energy range. For $\rho_D$, the minimum value of it appears near $t'=0.1$. Our findings regarding the enhancement of $d$-wave pairing for $t'<0$ are consistent with numerous studies indicating that negative $t'$ enhances $d$-wave superconductivity in the single-band Hubbard model~\cite{Jia2011, Sakakibara2010, Jiang2019}.\\

% Fig.4
%{\color{red} ------------CP pause here--------------}\\
%\noindent
%\textbf{Effect of Hubbard repulsion $U$ at half-filling}
%\\

\begin{figure}[htbp]
\centering
\includegraphics[width=0.9\linewidth]{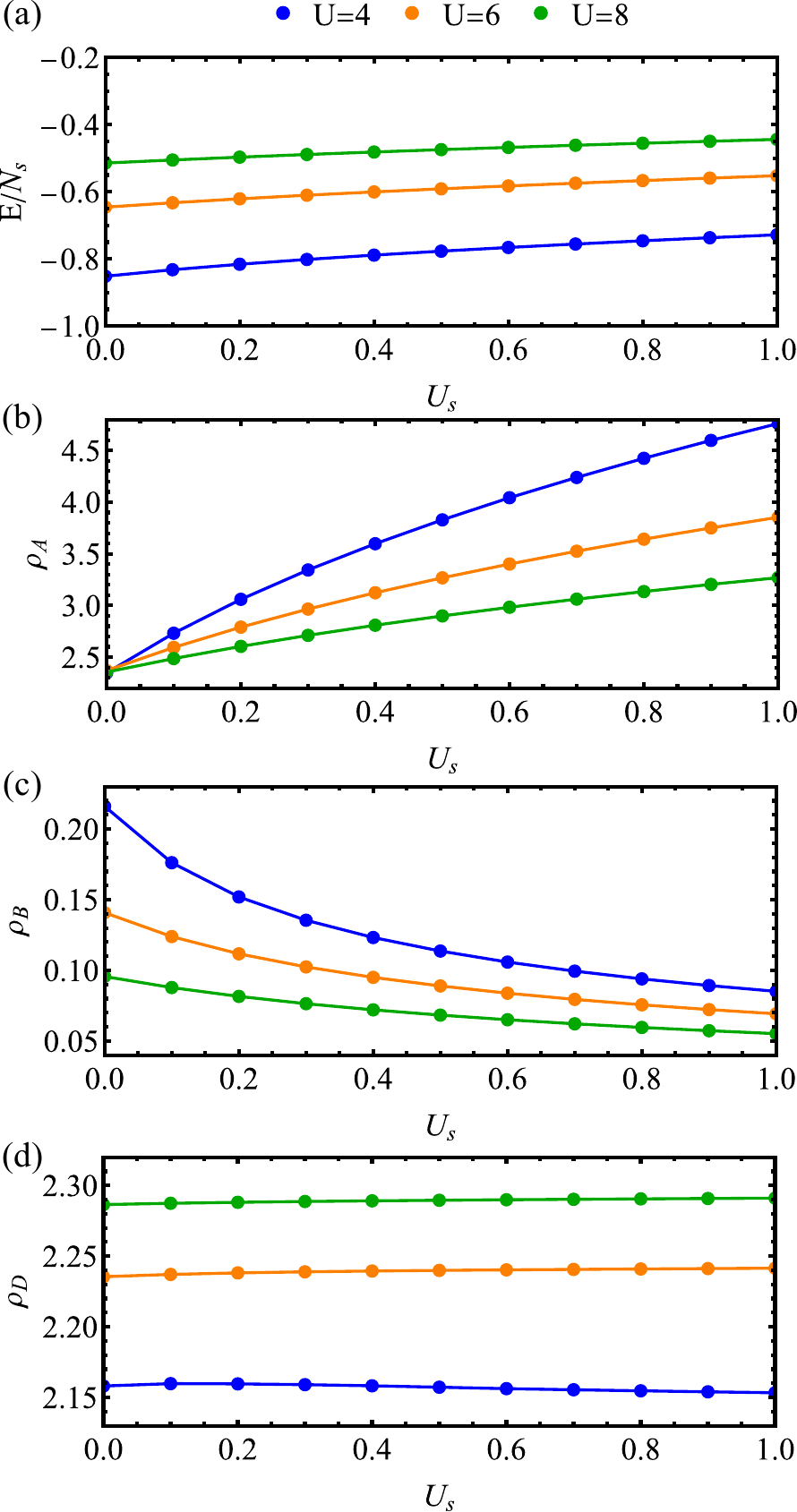}
\caption{(a) Ground state energy $E/N_s$, (b) extended $s$-wave pair density $\rho_A$, (c) $s$-wave pair density $\rho_B$ and (d) $d$-wave pair density $\rho_D$ versus strength $U_s$ of projection term $\hat{H}_{\text{proj}}$ under different Hubbard repulsion $U$ with $t'=0$ at half-filling and system's length $L_x=16$. Blue, orange and green colors correspond to $U=4,6,8$.}
\label{fig:result(U468t'0)}
\end{figure}

%\subsection{Effect of Hubbard repulsion $U$ at half-filling}
At last, we focus on how the value of Coulomb interaction $U$ affects the enhancement of extended $s$-wave pairing in the {\it order by production} scenario for the Hubbard model. Fig.~\ref{fig:result(U468t'0)} shows the calculated ground state energy per site, $E/N_s$, alongside three kinds of pair densities as they evolve with increasing $U_s$ for different Hubbard repulsion $U =4,6,$ and $8$. The behavior of the ground state energy and the pair densities share the same qualitative characteristics. Fig. \ref{fig:result(U468t'0)} (a) and (d) show the ground state energy and the $d$-wave Cooper pair density $\rho_D$ are raised by larger $U$, which is consistent with the general characteristic of the Hubbard model. Fig. \ref{fig:result(U468t'0)} (b) and (c) show both $\rho_A$, $\rho_B$ and their slopes $|\partial \rho_A/\partial U_s|$, $|\partial \rho_B/\partial U_s|$ increase simultaneously as Hubbard repulsion $U$ decreases from $8$ to $4$, which indicates that the enhancement (suppression) of $\rho_A$ $(\rho_B)$ with growing $U_s$ becomes more remarkable. 

The $U$ dependence can be understood in the following way: The projection terms $\hat{c}^{\dagger}_{i\uparrow}\hat{c}^{\dagger}_{i\downarrow}\hat{c}_{j\downarrow}\hat{c}_{j\uparrow}$ represent the hopping of a $s$-wave Cooper pair, which annihilates two electrons at a double-occupancy site $j$ and creates them at an empty site $i$. %However, the system with a large on-site Hubbard repulsion $U$ at half-filling is more likely to host one electron per site $\ket{\uparrow}$ and $\ket{\downarrow}$ in the antiferromagnetic Mott insulating phase, instead of the double-occupancy $\ket{\uparrow\downarrow}$ or an empty site $\ket{0}$. In other words, a small $U$ enlarges both the densities of the double-occupancy $\ket{\uparrow\downarrow}$ and the empty site $\ket{0}$, which leads to the effect of projection strength $U_s$ becomes more significant. 
However, in a system with a large on-site Hubbard repulsion $U$ at half-filling, there is a greater likelihood of hosting one electron per site $\ket{\uparrow}$ and $\ket{\downarrow}$ in the antiferromagnetic Mott insulating phase, rather than double-occupancy $\ket{\uparrow\downarrow}$ or an empty site $\ket{0}$. In other words, a smaller $U$ results in an increase in the densities of both double-occupancy $\ket{\uparrow\downarrow}$ and empty site $\ket{0}$, thereby amplifying the effect of the projection with strength $U_s$. \\

\noindent
\textbf{\emph{Order by projection} at light doping}
\\

In this section, we investigate the effect of hole-doping on the \emph{order by projection} mechanism when the Hubbard repulsion is finite. To facilitate comparison with the half-filling case, we apply hole-doping levels of $\delta=1/32$ and $\delta=1/16$ in the calculations at $U=8$ with a system length of $L_x=16$. The results for the ground-state energy per site, $E/N_s$, and the variation of three correlation functions with increasing $U_s$ are shown in Fig.~\ref{fig:result(U8t'0dop)}. Additional results for a longer system are provided in the supplementary material \cite{splm}.

\begin{figure}[ht!]
\centering
\includegraphics[width=0.9\linewidth]{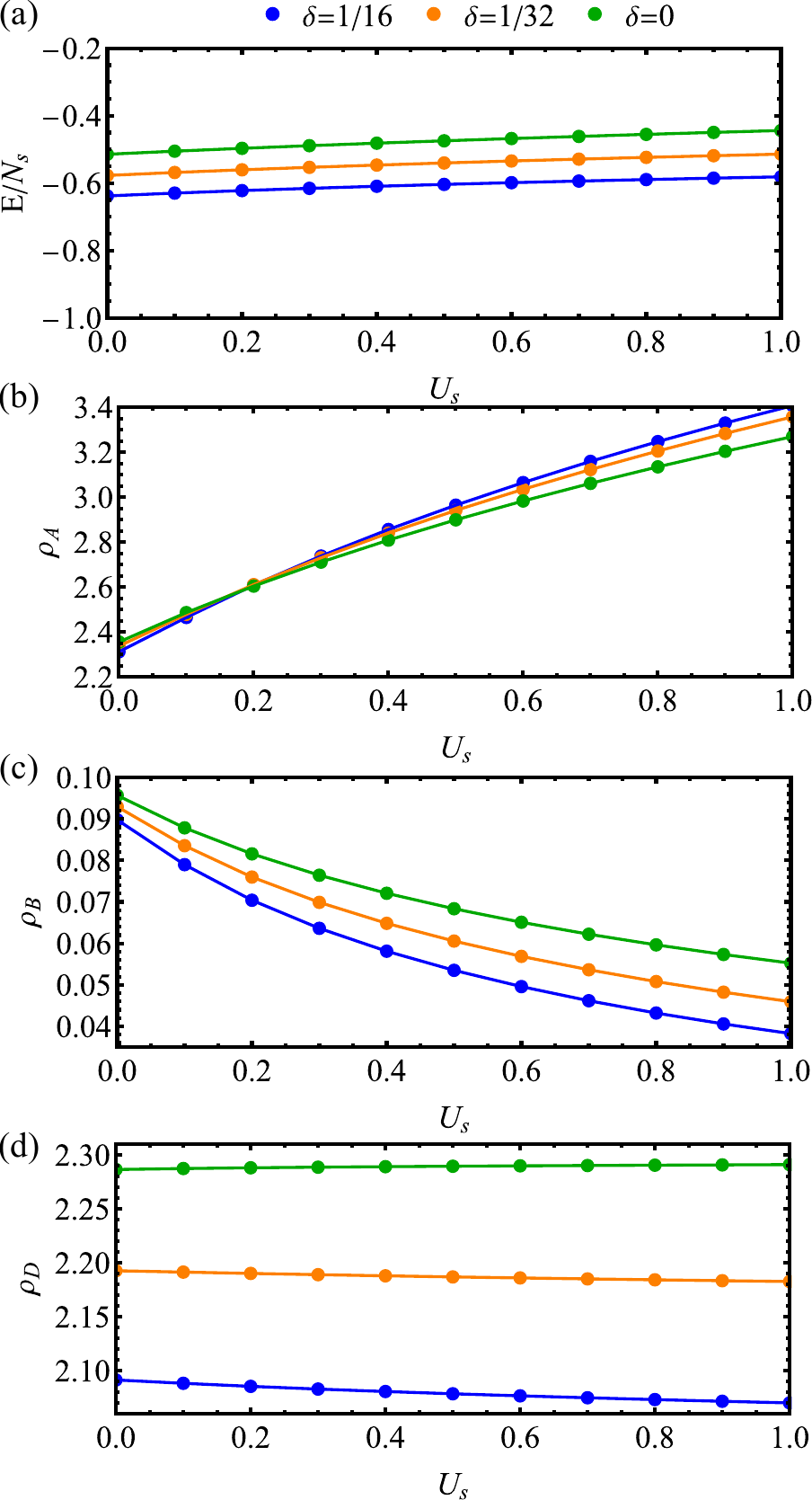}
\caption{(a) Ground state energy $E/N_s$, (b) extended $s$-wave pair density $\rho_A$, (c) $s$-wave pair density $\rho_B$ and (d) $d$-wave pair density $\rho_D$ versus strength $U_s$ of projection term $\hat{H}_{\text{proj}}$ under different hole doping $\delta$ with $U=8$ and $t'=0$ at system's length $L_x=16$. Different colors correspond to specific doping concentration.}
\label{fig:result(U8t'0dop)}
\end{figure}

In contrast to the case at half-filling, the ground-state energy decreases with hole-doping, while the effect of $U_s$ on the enhancement (suppression) of $\rho_A$ $(\rho_B)$ with increasing $U_s$, i.e. $|\partial \rho_A/\partial U_s|$ and $|\partial \rho_B/\partial U_s|$, becomes more remarkable. Notice that the projection term $\hat{H}_{\text{proj}}$ contains the $s$-wave Cooper pair hopping between any two sites. When the system slightly moves away from half-filling, the density of electrons decreases at hole-doping, leading to a decrease $|\Delta\rho_{\text{double}}|$ in the density of the double-occupancy $\ket{\uparrow\downarrow}$ and increase $|\Delta\rho_{\text{empty}}|$ in the density of empty site $\ket{0}$. However, under a finite positive on-site $U$, the hole-doping caused increasing in the density of empty site $|\Delta\rho_{\text{empty}}|$ is much larger than its reduction in the density of the double-occupancy $|\Delta\rho_{\text{double}}|$ (it is about $|\Delta\rho_{\text{empty}}|/|\Delta\rho_{\text{double}}|>10$, as shown in supplementary material~\cite{splm} Fig. S4 and S5). In other words, a little reduction in number of double occupancy sites, but each pair of electrons on the double occupancy sites now has the potential to hop to much more choices of empty sites in the hole-doped case. Thus, the \emph{order by projection} effect is also enhanced by hole-doping, as shown in the enhanced extended $s$-wave density $\rho_A$ for $U_s \sim 1$ and a larger slope of $\rho_A$ as a function of $U_s$ at increased hole-doping concentration.\\ %However, the variation in the magnitude of $\rho_A$ and $\rho_B$ is different because the hole-doping also has a significant effect on the normal hopping terms. {\color{red} Some information may be missing here, and I don't understand why doping could modify normal hopping and how does this relates to the magnitude of $\rho_A$ and $\rho_B$.}

\vspace{8mm}
\noindent
\MakeUppercase{\textbf{Discussion}}
\\

\begin{figure}[htbp]
\centering
\includegraphics[width=0.9\linewidth]{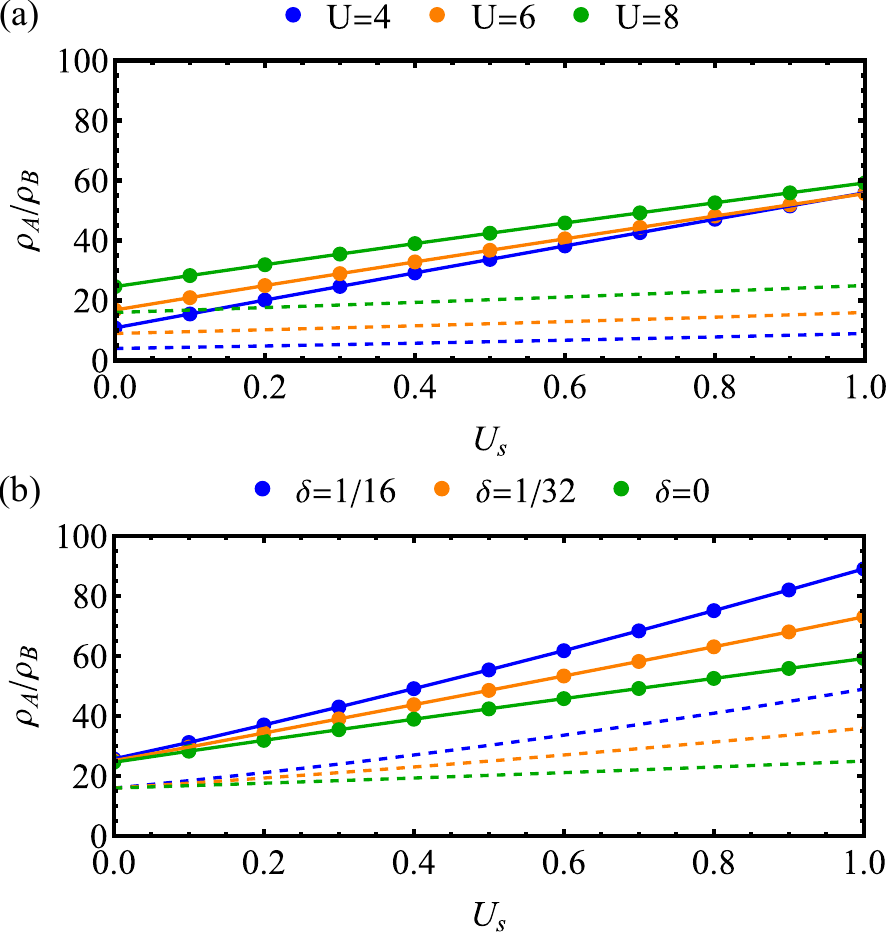}
\caption{The ratio between extended $s$-wave pair density and $s$-wave pair density $\rho_A/\rho_B$ versus strength $U_s$ of projection term $H_{\text{proj}}$ under (a) different Hubbard repulsion $U$ with $t'=0$ at half-filling and (b) different hole doping $\delta$ with $U=8$ and $t'=0$. The dashed lines show $(U_s(N_s-N_e+2)+U)^2/4$ for each $U$ or doping case, where $N_s$ is the system size and $N_e$ is the number of electrons.}
\label{fig:result(AoverB)}
\end{figure}

In this paper, we have studied the influence of projecting out the $s$-wave Cooper pairs on a Fermi Hubbard model with finite Hubbard repulsion $U$ by DMRG method. %\lsy{Fig.\ref{fig:result(U8t'0)} explores the behavior of extended $s$-wave, $s$-wave, and $d$-wave pair-density matrices with increasing projection strength $U_s$.}
As the projection strength $U_s$ increases, the $s$-wave Cooper pair density $\rho_B$ decreases while the extended $s$-wave Cooper pair density $\rho_A$ is enhanced remarkably even though it does not appear in the Hamiltonian, which provides numerical evidence to the \emph{order by projection} mechanism with finite on-site repulsion $U$. In previous work~\cite{orderbyprojection1}, a lower bound of the ratio between $\rho_A$ and $\rho_B$ is given as $\rho_A/\rho_B \geq (U_s(N_s-N_e+2)+U-2\mu)^2/4$, where $N_e$ is the number of electrons and $\mu$ is the chemical potential. As shown in Fig.~\ref{fig:result(AoverB)}, our DMRG calculation of $\rho_A/\rho_B$ remains above $(U_s(N_s-N_e+2)+U)^2/4$ for all doping levels and Hubbard $U$ values studied, which makes the above inequality automatically hold. Also, the scaling behavior of $\rho_A$ and the ratio $R_A$ shown in Fig.~\ref{fig:result(U8t'0)R} indicate that the \emph{order by projection} leads to the appearance of quasi-long-range order and the condensation of extended $s$-wave Cooper pairs. In addition, the extended $s$-wave and $d$-wave ordering could be altered by $t'$ as evidenced by Fig.~\ref{fig:result(U8t')}, whereas the $s$-wave channel is almost unchanged.\\

Interestingly, we found a competition between Hubbard repulsion $U$ and the hole-doping level away from half-filling in the enhancement of the \emph{order by projection} effect. The presence of a strong on-site Hubbard repulsion $U$ in the system suppresses the distribution of the double occupancy $\ket{\uparrow\downarrow}$ and the empty site $\ket{0}$, thus making the \emph{order by projection} effect less pronounced. Meanwhile, in the case of $U=0$ near half-filling~\cite{orderbyprojectionexact}, the exact solution in the thermodynamic limit indicates that $\rho_A \sim 1/\delta$, which shows that hole doping $\delta$ decreases $\rho_A$, as well as its respective rate of change with respect to $U_s$: $|\partial\rho_A/\partial U_s|$. However, DMRG results at finite $U$ provide positive evidence contrary to this analytical analysis for $U/t>0$: we found that $|\partial\rho_A/\partial U_s|$ is enhanced by hole doping.\\

%\sr{I feel at his location you should add a small discussion of Fig.4 and the effect of t' on d-wave pairing. I also wonder if it is worth making a small catalog of all the main figures in the paper, with a brief commentary. As an example you could say "Fig. 4 explores the role of t' on the different pairing channels. We see that the d-wave ordering is encouraged by this term, whereas the other channels are.."}

Besides the projection of the $s$-wave Cooper pair, the \emph{order by projection} mechanism is reported to be valid for a general form $\hat{H}_{\text{proj}}=U_s \hat{B}^{\dagger}\hat{B}$ with $\hat{B}=\sum_{\vec{k}}e^{i\phi_{\vec{k}}}\hat{c}_{-\vec{k}\downarrow}\hat{c}_{\vec{k}\uparrow}$ at $U=0$. It will be an interesting addition to investigate the projection out of other Cooper pair channels, such as $d$-wave pairs in the presence of Hubbard repulsion $U$, as well as on other lattices, such as triangular and honeycomb ones. Moreover, the pair-hopping terms are more likely to be dependent on the distance between two sites, so it is natural to ask whether the \emph{order by projection} mechanism still works for this case. In the simplest case, that the projection term $\hat{H}_{\text{proj}}$ only contains the interaction between the first and second nearest neighbor sites (more details are provided in supplementary material~\cite{splm}), this effect on $\rho_A$ and $\rho_B$ becomes weak but it still works.  These open questions will merit further research in the future.\\

\vspace{8mm}
\noindent
\MakeUppercase{\textbf{Methods}}
\\

Our DMRG calculations are performed under the conservation of the good quantum number of total electrons with equal spin up and down, ensuring that $\sum_{N_s} \langle \hat{S}^z_i \rangle = 0$ for both half-filling and hole doped cases. We maintain bond dimensions between 5000 and 7000 to achieve results with a satisfactory truncation error of $\epsilon \sim 10^{-6}$. The convergence quality of our results with increasing bond dimension has been investigated and is shown in the supplementary material~\cite{splm}.

\vspace{4mm}
\noindent
\MakeUppercase{\textbf{Code Availability}}\\
{\small
The codes and scripts used for the numerical calculations reported in this paper are available from the first author (S.L.) upon request.
}\\

\noindent
\MakeUppercase{\textbf{Data Availability}}\\
{\small
The data analyzed in the present study is available from the first author (S.L.) upon reasonable request.
}

\bibliographystyle{nature}
\bibliography{cite}

\vspace{4mm}
\noindent
\uppercase{\textbf{Acknowledgements}}\\

This work is supported by the U.S. Department of Energy, Office of Science, Basic Energy Sciences under Award No. DE-SC0022216. DMRG calculations were performed using the \emph{ITensor} library~\cite{itensor}.

\vspace{4mm}
\noindent
\uppercase{\textbf{Author contributions}}\\

C.J. and B.S.S conceived and supervised the work. S.L. and C.P. performed DMRG calculations. S.L. analyzed the data. Y.Y. set up the calculation environment of HiPerGator. S.L., C.J. C.P., and B.S.S. wrote the manuscript, with input from all authors. 

\vspace{4mm}
\noindent
\uppercase{\textbf{Competing interests}}\\

The authors declare no competing interests.

\vspace{4mm}
\noindent
\uppercase{\textbf{Additional Information}}\\
{\small \textbf{Supplementary Information} The online version contains supplementary material available at \ldots\\

\noindent
\textbf{Correspondence} should be addressed to C.J. Request for materials should be directed to S.L..
}

\onecolumngrid
\newpage

\noindent
\textbf{Supplementary Materials for ``Order by projection in single-band Hubbard model: a DMRG study"}\\

\renewcommand{\theequation}{S\arabic{equation}}
\renewcommand{\thefigure}{S\arabic{figure}}
\renewcommand{\bibnumfmt}[1]{[S#1]}
\renewcommand{\citenumfont}[1]{S#1}

\section{Pair density matrices in the region of negative $U_s$}

\begin{figure}[htbp!]
\centering
\includegraphics[width=0.9\linewidth]{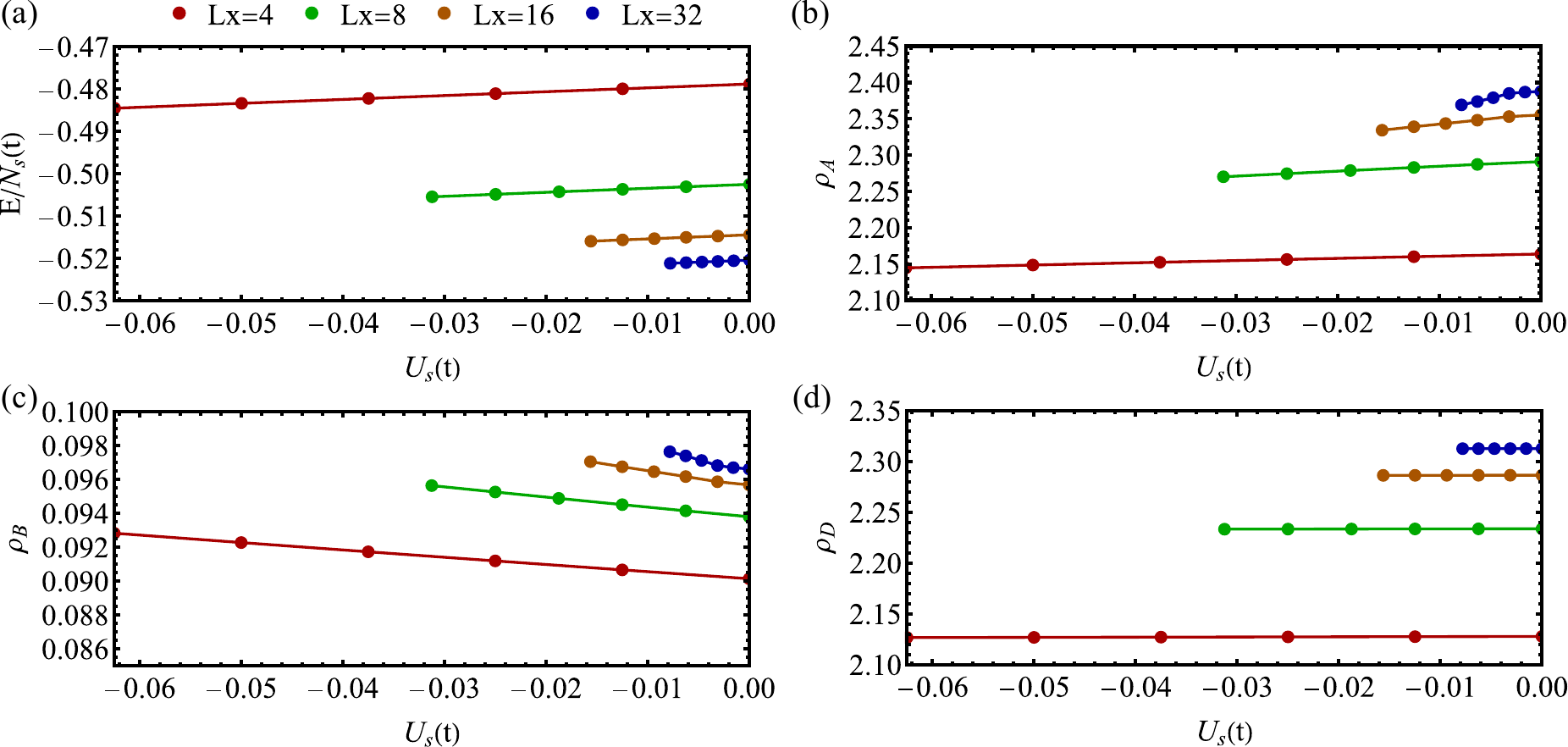}
\caption{In the region of $U_s\leq0$, (a) ground state energy $E/N_s$, (b) extended $s$-wave pair density $\rho_A$, (c) $s$-wave pair density $\rho_B$ and (d) $d$-wave pair density $\rho_D$ versus strength $U_s$ of projection term $\hat{H}_{\text{proj}}$ under $U=8t$, $t'=0$. Red, green, brown, and blue colors correspond to the system's length $L_x=4,8,16,32$.}
\label{fig:result(U8t'0)n}
\end{figure}

In the case of $U/t=8$ and $t'=0$ at half-filling, as a comparison, the ground state energy per site $E/N_s$ and the pair densities for extended $s$-wave pair density $\rho_A$, $s$-wave pair density $\rho_B$, and $d$-wave pair density $\rho_D$ as functions of negative $U_s$ are shown in Fig.~\ref{fig:result(U8t'0)n}, where $U_s$ is the range of $-t/N_s<U_s<0$. With an increase in $|U_s|$, $\rho_B$ is enhanced while $\rho_A$ is suppressed.

\section{Ratio between the largest and second-largest eigenvalues of the pair-density matrix}

\begin{figure}[htbp!]
\centering
\includegraphics[width=0.9\linewidth]{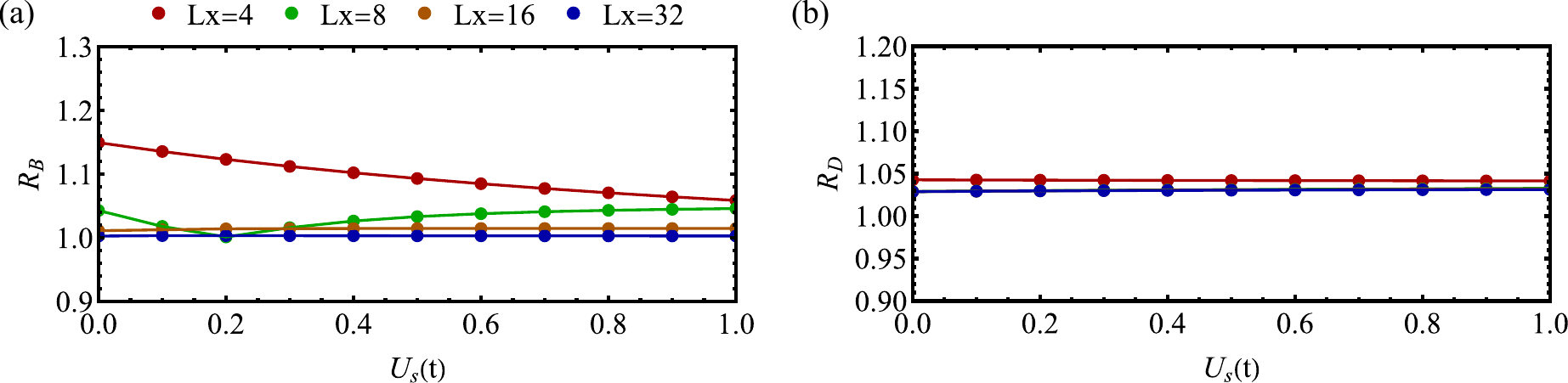}
\caption{The ratio between the largest and second-largest eigenvalues of the (a) $s$-wave pair-density matrix $R_B$ and (b) $d$-wave pair-density matrix $R_D$ versus $U_s$ in the case of $U/t=8$ and $t'=0$ at half-filling for different length $L_x$. Red, green, brown, and blue colors correspond to the system's length $L_x=4,8,16,32$.}
\label{fig:result(U8t'0)RBD}
\end{figure}

In the case of $U/t=8$ and $t'=0$ at half-filling, the ratios $R_B$ and $R_D$ as functions of $U_s$ are shown in Fig. \ref{fig:result(U8t'0)RBD}. They are almost unchanged for large system size $L_x\geq16$.

\newpage
\section{Hole doping near half-filling at system length $\mathbf{L_x=20}$}

In the case of $U/t=8$ and $t'=0$ at system length $L_x=20$, the results of the ground-state energy per site $E/N_s$ and three correlation functions varying with increasing $U_s$ are shown in Fig.~\ref{fig:result(U8t'0dop)}. 

\begin{figure}[htbp!]
\centering
\includegraphics[width=0.9\linewidth]{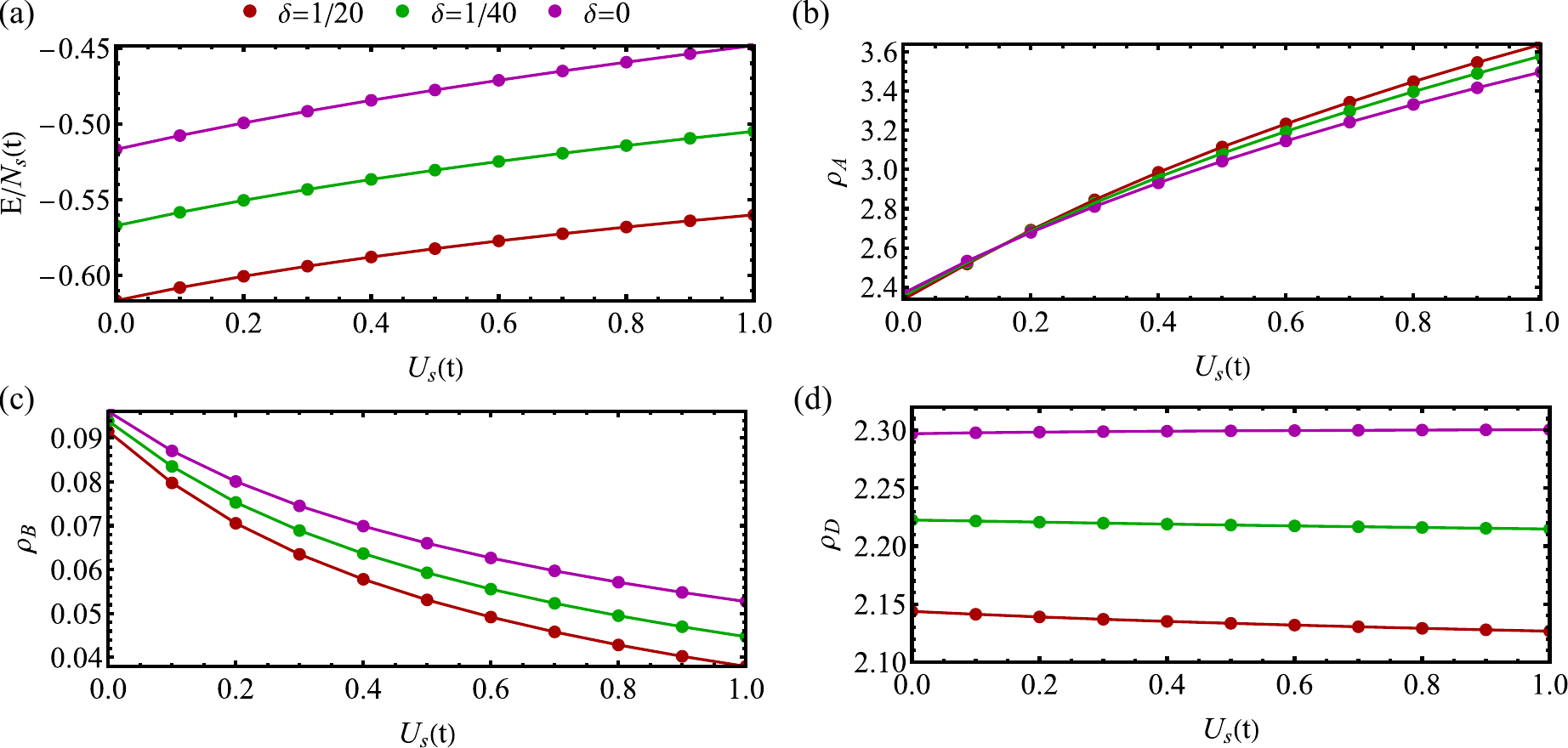}
\caption{(a) Ground state energy $E/N_s$, (b) extended $s$-wave pair density $\rho_A$, (c) $s$-wave pair density $\rho_B$ and (d) $d$-wave pair density $\rho_D$ versus strength $U_s$ of projection term $\hat{H}_{\text{proj}}$ under different hole doping $\delta$ with $U=8t$ and $t'=0$ at system's length $L_x=20$. Red, green, and purple colors correspond to $\delta=1/20,1/40,0$.}
\label{fig:result(U8t'0dop)}
\end{figure}

\section{The density of double occupancy and empty sites}
\begin{figure}[ht!]
\centering
\includegraphics[width=0.5\linewidth]{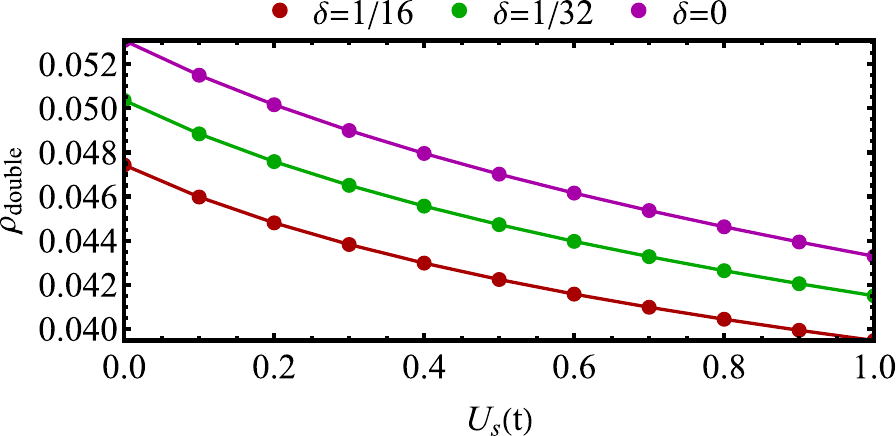}
\caption{The density of double occupancy $\rho_{\text{double}}$ versus strength $U_s$ of projection term $\hat{H}_{\text{proj}}$ under different hole doping $\delta$ with $U=8t$, $t'=0$ at system length $L_x=16$. Red, green, and purple colors correspond to $\delta=1/16,1/32,0$.}
\label{fig:result(U8t'0double)}
\end{figure}

\begin{figure}[ht!]
\centering
\includegraphics[width=0.9\linewidth]{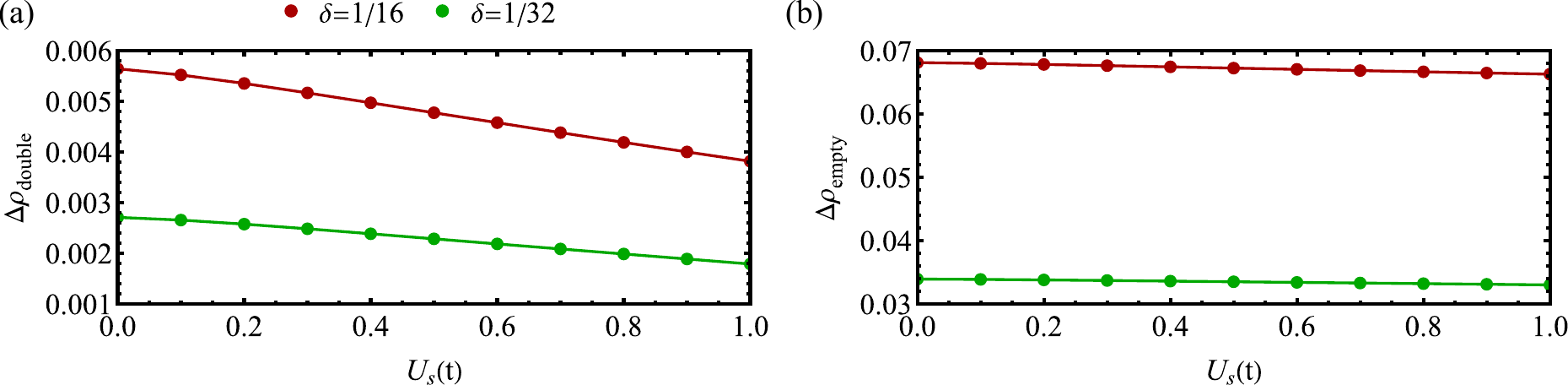}
\caption{(a) The change in the density of double occupancy $\Delta\rho_{\text{double}}$ and (b) the change in the density of empty sites $\Delta\rho_{\text{empty}}$ versus strength $U_s$ of projection term $\hat{H}_{\text{proj}}$ under different hole doping $\delta$ with $U=8t$, $t'=0$ at system length $L_x=16$. Red and green colors correspond to $\delta=1/16,1/32$.}
\label{fig:result(U8t'0changede)}
\end{figure}

In this section, we provide the results about the density of double occupancy and empty sites, which are defined by
\begin{equation}
    \rho_{\text{double}}=\frac{\langle \sum_i n_{i\uparrow}n_{i\downarrow}\rangle}{N_s},
\end{equation}
\begin{equation}
    \rho_{\text{empty}}=\frac{\langle \sum_i (1-n_{i\uparrow})(1-n_{i\downarrow})\rangle}{N_s}=\rho_{\text{double}}+\delta,
\end{equation}
where $\delta$ is the ratio of hole-doping. In the case of $U/t=8$ and $t'=0$ at system length $L_x=16$, the plot of $\rho_{\text{double}}$ as a function of $U_s$ for $\delta=1/16,1/32,0$ is shown in Fig.~\ref{fig:result(U8t'0double)}. After applying hole-doping, the change in the density of double occupancy and empty sites are $\Delta\rho_{\text{double}}$ and $\Delta\rho_{\text{empty}}$, which are shown in Fig.~\ref{fig:result(U8t'0changede)}. In the range of $U_s>0$, $\Delta\rho_{\text{empty}}>10*\Delta\rho_{\text{double}}$, which makes the effect of projection term become more remarkable. 

\section{The cases of short-range interactions}

In this section, we provide some examples of results from the projection terms that only contain short-range interactions. The full projection term considered in the main text is:
\begin{equation}
    \hat{H}_{\text{proj}}=U_s \hat{B}^{\dagger}\hat{B}=U_s \sum_{i,j}\hat{c}_{i\uparrow}^{\dagger}\hat{c}_{i\downarrow}^{\dagger}\hat{c}_{j\downarrow}\hat{c}_{j\uparrow}.
\end{equation}
Here we consider two cases: the projection term with the first nearest-neighbor interaction, and another one with both the first and second nearest-neighbor interactions, which are given by 

\begin{equation}
    \hat{H}_{\text{proj}}^{\langle1\rangle}=U_s \sum_{\langle ij\rangle_1}\hat{c}_{i\uparrow}^{\dagger}\hat{c}_{i\downarrow}^{\dagger}\hat{c}_{j\downarrow}\hat{c}_{j\uparrow},~\hat{H}_{\text{proj}}^{\langle2\rangle}=U_s \sum_{\langle ij\rangle_1}\hat{c}_{i\uparrow}^{\dagger}\hat{c}_{i\downarrow}^{\dagger}\hat{c}_{j\downarrow}\hat{c}_{j\uparrow}+U_s \sum_{\langle ij\rangle_2}\hat{c}_{i\uparrow}^{\dagger}\hat{c}_{i\downarrow}^{\dagger}\hat{c}_{j\downarrow}\hat{c}_{j\uparrow}.
\end{equation}

\begin{figure}[ht!]
\centering
\includegraphics[width=0.9\linewidth]{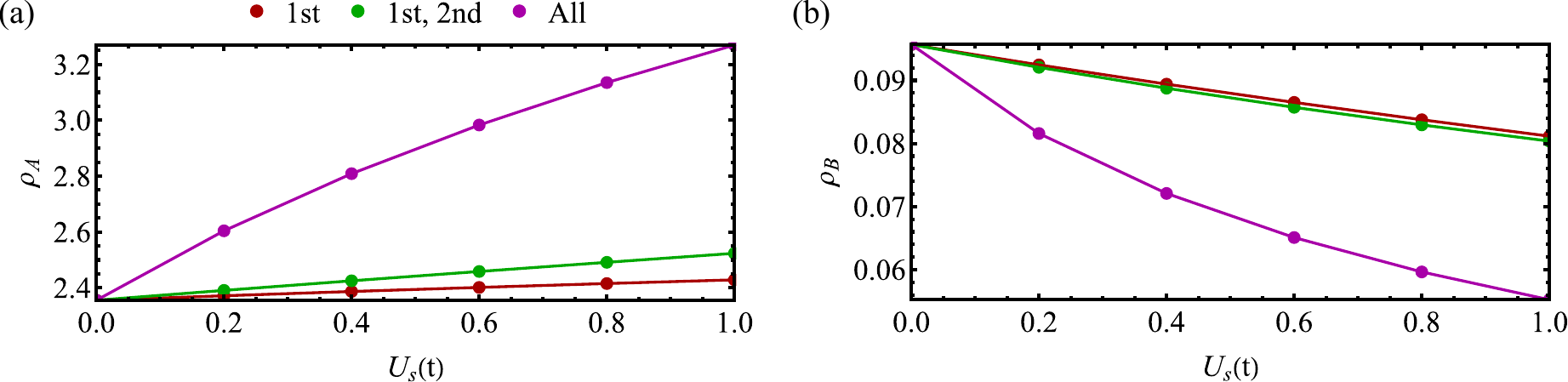}
\caption{(a) Extended $s$-wave pair density $\rho_A$ and (b) $s$-wave pair density $\rho_B$ versus strength $U_s$ of projection term $\hat{H}_{\text{proj}}$ under $U=8t$, $t'=0$ at system length $L_x=16$. Red, green, and purple colors correspond to the system with the first neighbor pairing hoping, the first and second neighbor pairing hoping, and pairing hoping between all neighbors.}
\label{fig:result(U8t'0SR)}
\end{figure}

Under $U=8t$, $t'=0$ at system length $L_x=16$, the pair densities for extended $s$-wave pair density $\rho_A$ and $s$-wave pair density $\rho_B$ as functions of negative $U_s$ are shown in Fig.~\ref{fig:result(U8t'0SR)}.

\section{Convergence of DMRG calculation with increasing bond dimension}

\begin{figure}[htbp]
\centering
\includegraphics[width=0.9\linewidth]{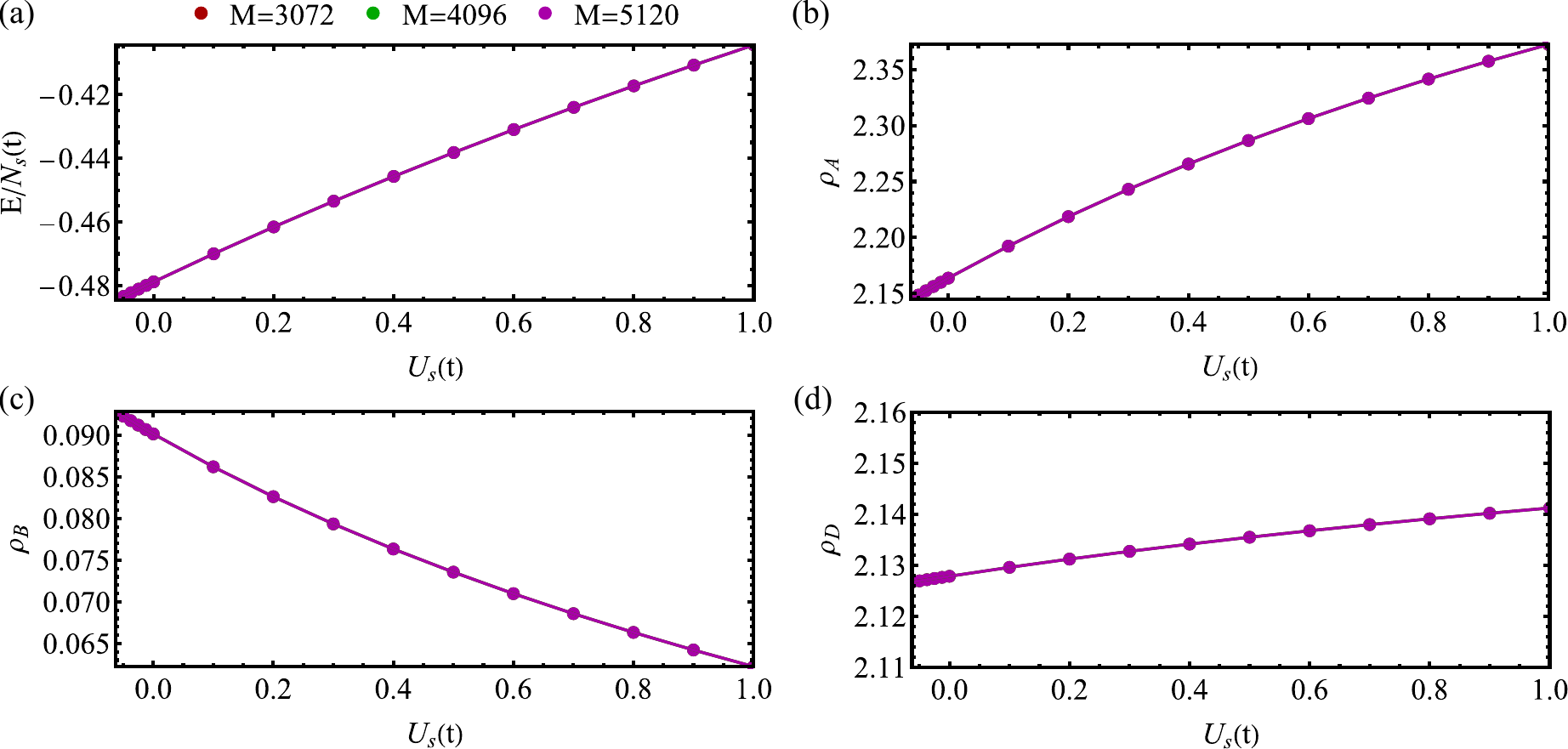}
\caption{Results of DMRG calculation at different bond dimensions: (a) ground state energy $E/N_s$, (b) extended $s$-wave pair density $\rho_A$, (c) $s$-wave pair density $\rho_B$ and (d) $d$-wave pair density $\rho_D$ versus strength $U_s$ of projection term $\hat{H}_{\text{proj}}$ under $U=8t$, $t'=0$ with the system's length $L_x=4$. Red, green, and purple colors correspond to the bond dimensions $M=3072, 4096, 5120$.}
\label{Fig:conv(U8t'0)4*4}
\end{figure}

\begin{figure}[htbp]
\centering
\includegraphics[width=0.9\linewidth]{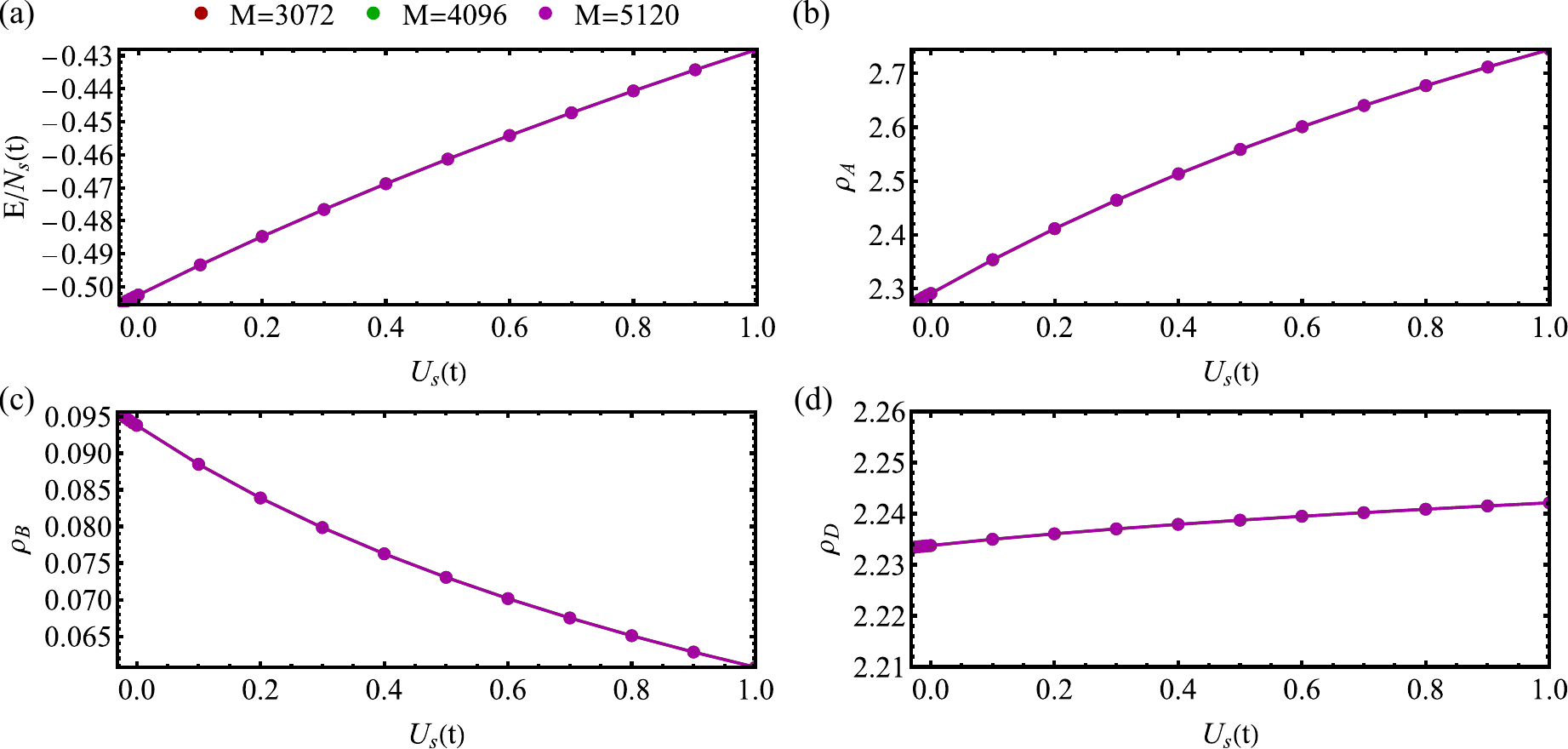}
\caption{Results of DMRG calculation at different bond dimensions: (a) ground state energy $E/N_s$, (b) extended $s$-wave pair density $\rho_A$, (c) $s$-wave pair density $\rho_B$ and (d) $d$-wave pair density $\rho_D$ versus strength $U_s$ of projection term $\hat{H}_{\text{proj}}$ under $U=8t$, $t'=0$ with the system's length $L_x=8$. Red, green, and purple colors correspond to the bond dimensions $M=3072, 4096, 5120$.}
\label{Fig:conv(U8t'0)8*4}
\end{figure}

\begin{figure}[htbp]
\centering
\includegraphics[width=0.9\linewidth]{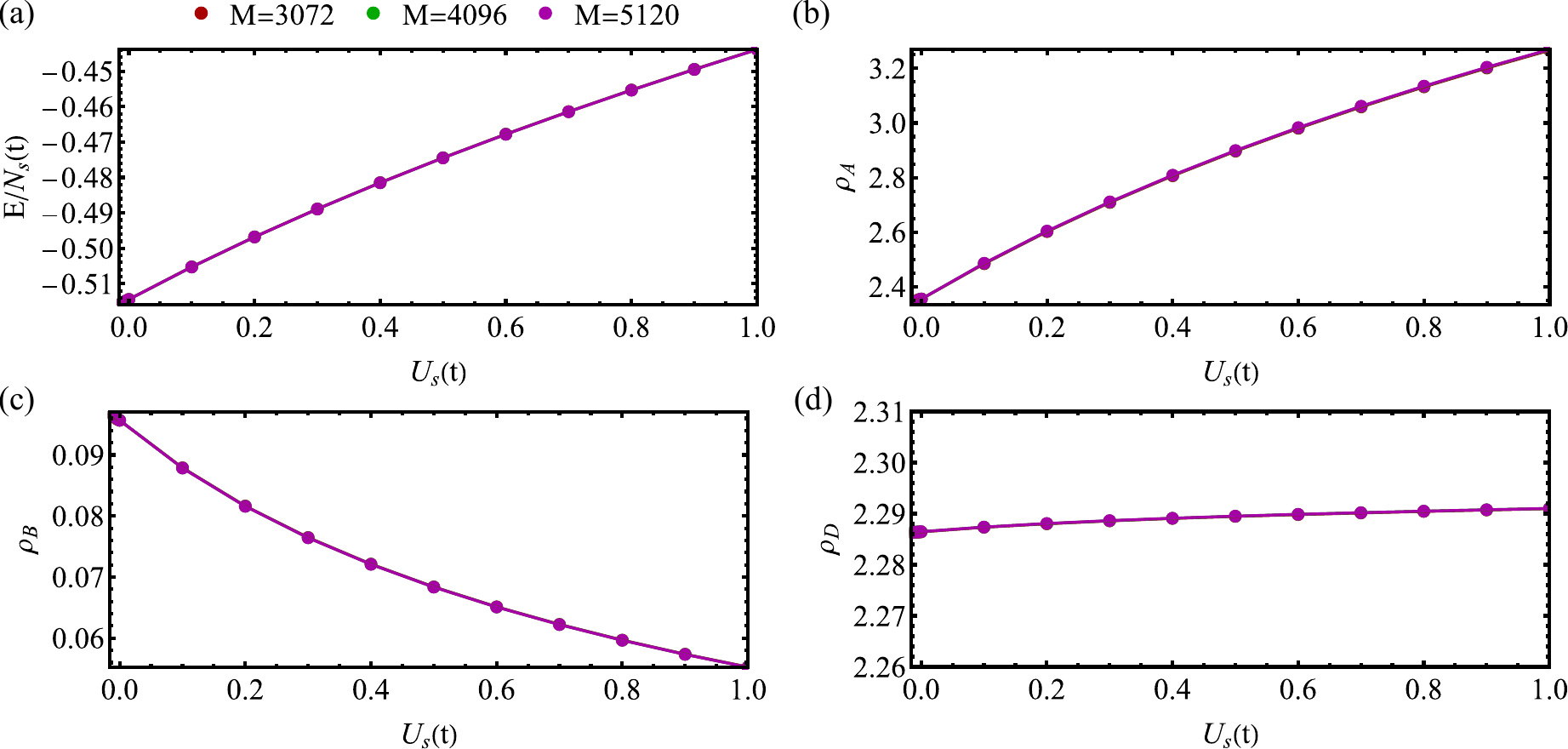}
\caption{Results of DMRG calculation at different bond dimensions: (a) ground state energy $E/N_s$, (b) extended $s$-wave pair density $\rho_A$, (c) $s$-wave pair density $\rho_B$ and (d) $d$-wave pair density $\rho_D$ versus strength $U_s$ of projection term $\hat{H}_{\text{proj}}$ under $U=8t$, $t'=0$ with the system's length $L_x=16$. Red, green, and purple colors correspond to the bond dimensions $M=3072, 4096, 5120$.}
\label{Fig:conv(U16t'0)4*4}
\end{figure}

\begin{figure}[htbp]
\centering
\includegraphics[width=0.9\linewidth]{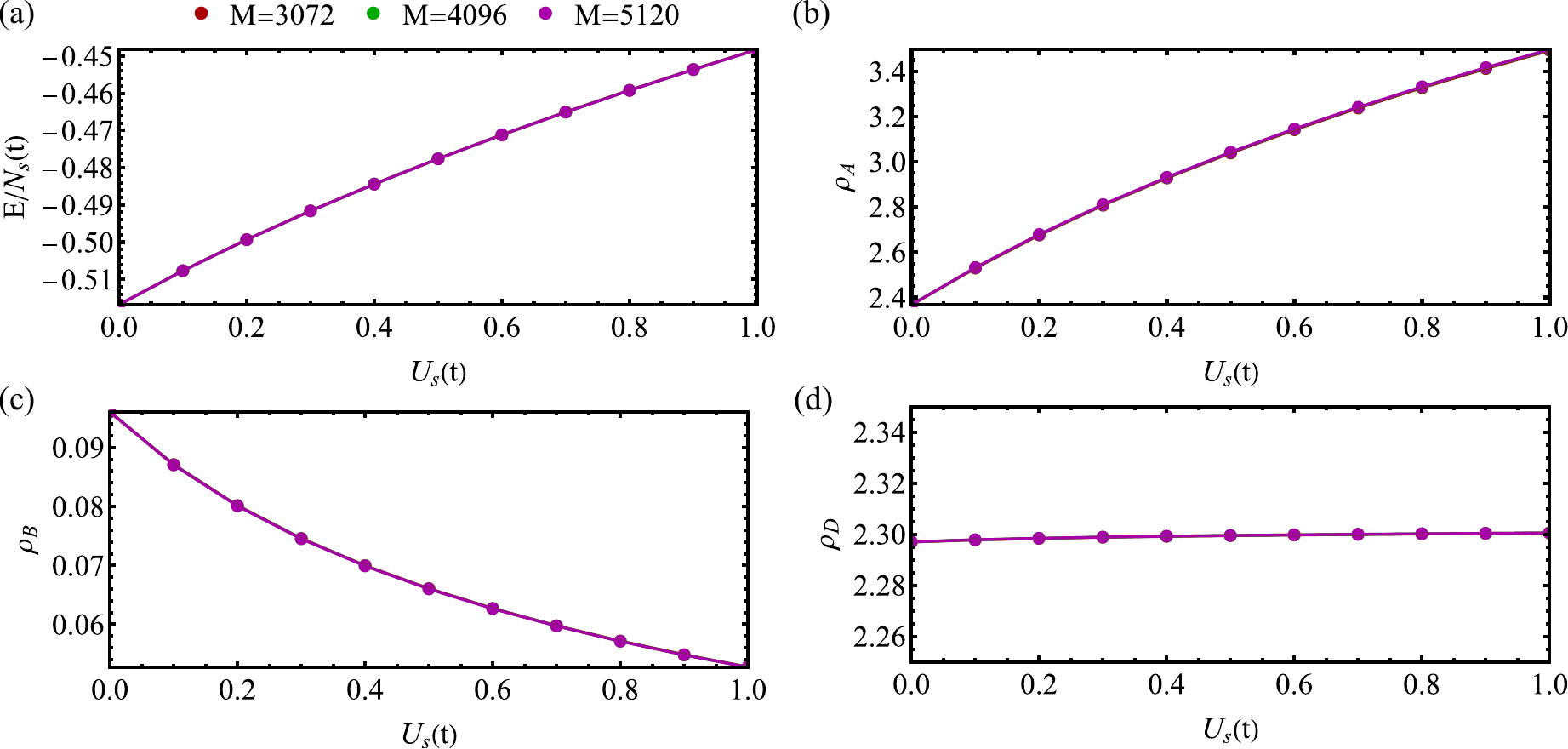}
\caption{Results of DMRG calculation at different bond dimensions: (a) ground state energy $E/N_s$, (b) extended $s$-wave pair density $\rho_A$, (c) $s$-wave pair density $\rho_B$ and (d) $d$-wave pair density $\rho_D$ versus strength $U_s$ of projection term $\hat{H}_{\text{proj}}$ under $U=8t$, $t'=0$ with the system's length $L_x=20$. Red, green, and purple colors correspond to the bond dimensions $M=3072, 4096, 5120$.}
\label{Fig:conv(U8t'0)20*4}
\end{figure}

\begin{figure}[htbp]
\centering
\includegraphics[width=0.9\linewidth]{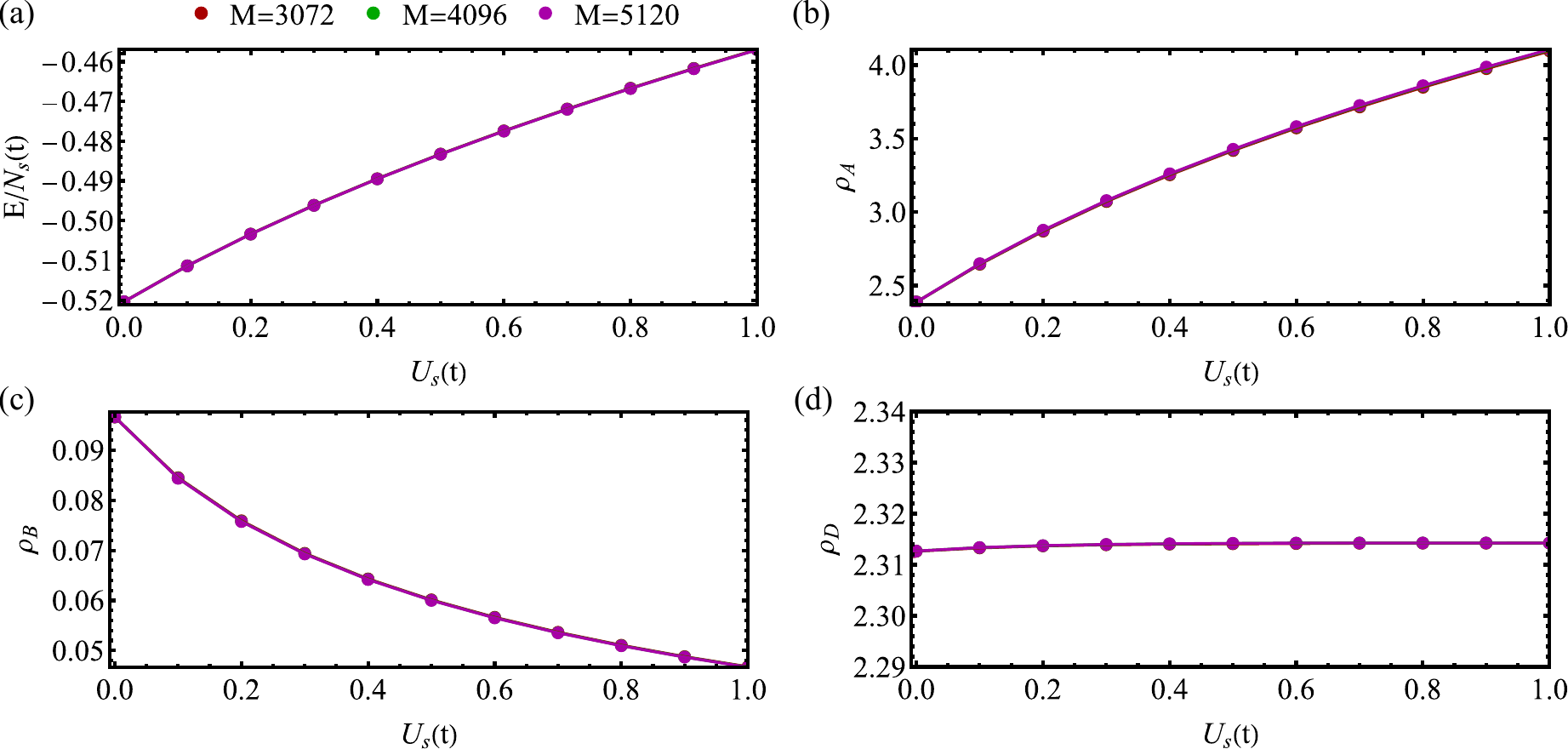}
\caption{Results of DMRG calculation at different bond dimensions: (a) ground state energy $E/N_s$, (b) extended $s$-wave pair density $\rho_A$, (c) $s$-wave pair density $\rho_B$ and (d) $d$-wave pair density $\rho_D$ versus strength $U_s$ of projection term $\hat{H}_{\text{proj}}$ under $U=8t$, $t'=0$ with the system's length $L_x=32$. Red, green, and purple colors correspond to the bond dimensions $M=3072, 4096, 5120$.}
\label{Fig:conv(U8t'0)32*4}
\end{figure}

\begin{figure}[htbp]
\centering
\includegraphics[width=0.9\linewidth]{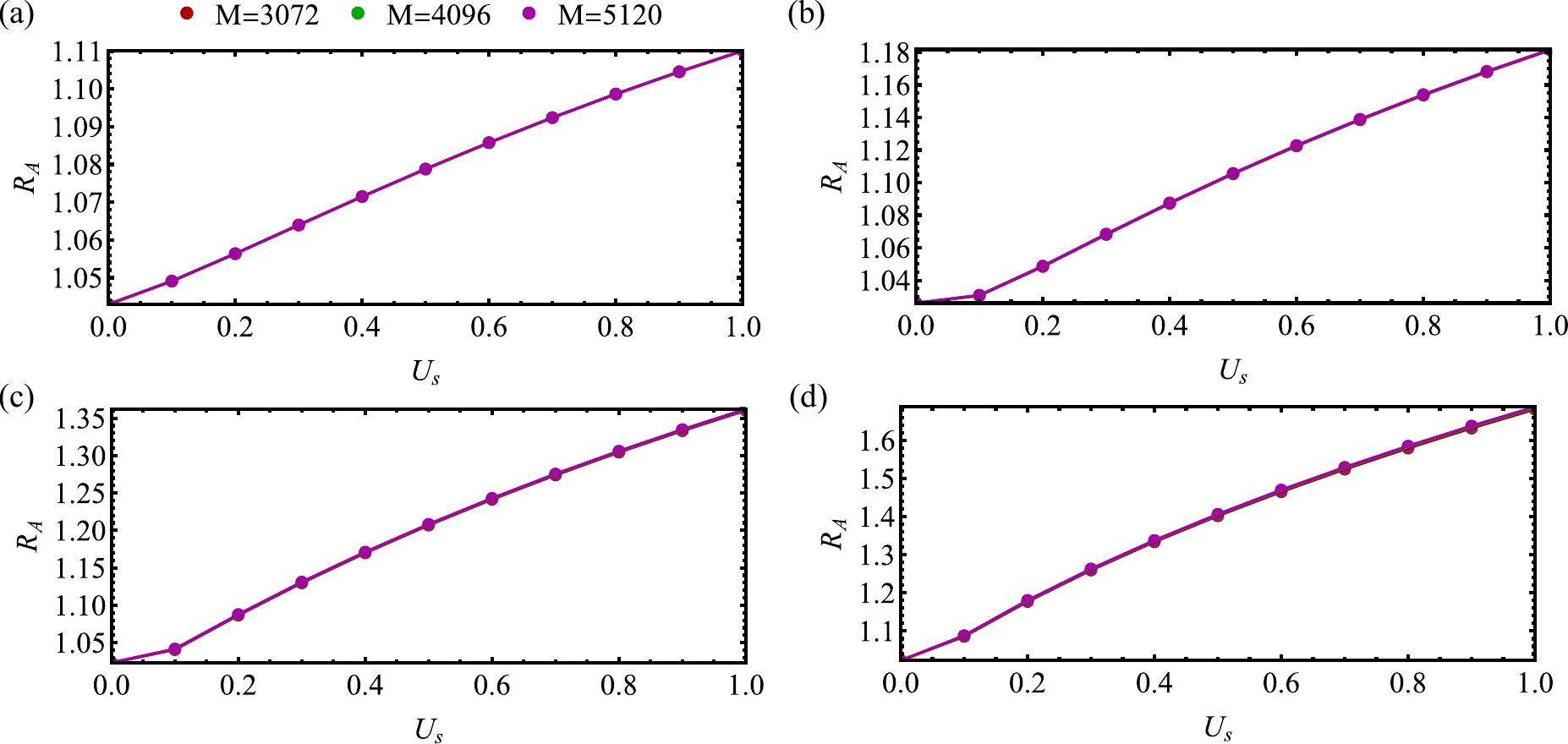}
\caption{Results of DMRG calculation at different bond dimensions: the ratio $R_A$ between the largest and second-largest eigenvalues of the extended s-wave pair-density matrix versus $U_s$ in the case of $U/t=8$ and $t'=0$ at half-filling for (a) $L_x=4$, (b) $L_x=8$, (c) $L_x=16$, and (d) $L_x=32$. Red, green, and purple colors correspond to the bond dimensions $M=3072, 4096, 5120$.}
\label{Fig:conv(RA)16*4}
\end{figure}

\begin{figure}[htbp]
\centering
\includegraphics[width=0.9\linewidth]{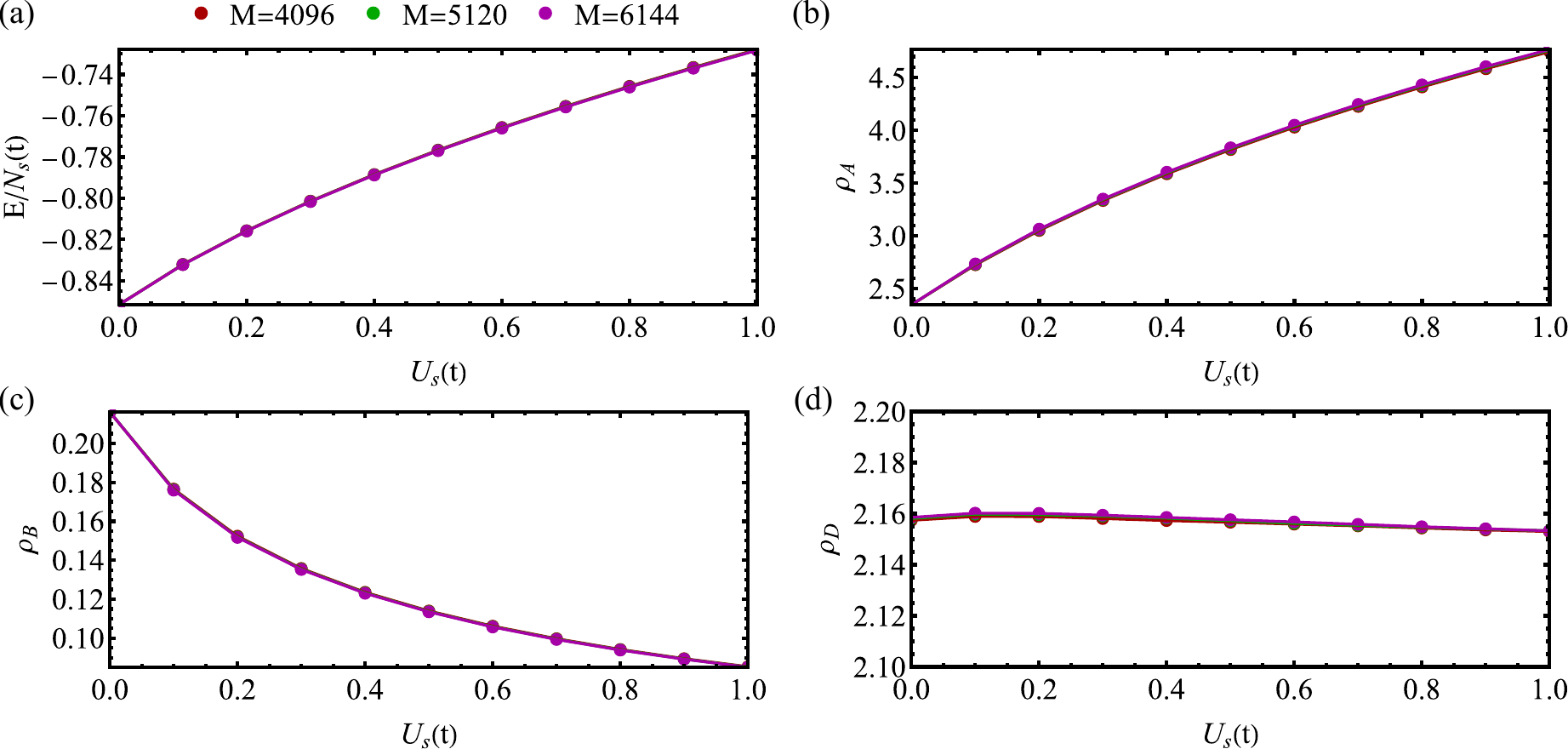}
\caption{Results of DMRG calculation at different bond dimensions: (a) ground state energy $E/N_s$, (b) extended $s$-wave pair density $\rho_A$, (c) $s$-wave pair density $\rho_B$ and (d) $d$-wave pair density $\rho_D$ versus strength $U_s$ of projection term $\hat{H}_{\text{proj}}$ under $U=4t$, $t'=0$ with the system's length $L_x=16$. Red, green, and purple colors correspond to the bond dimensions $M=4096, 5120, 6144$.}
\label{Fig:conv(U4t'0)16*4}
\end{figure}

\begin{figure}[htbp]
\centering
\includegraphics[width=0.9\linewidth]{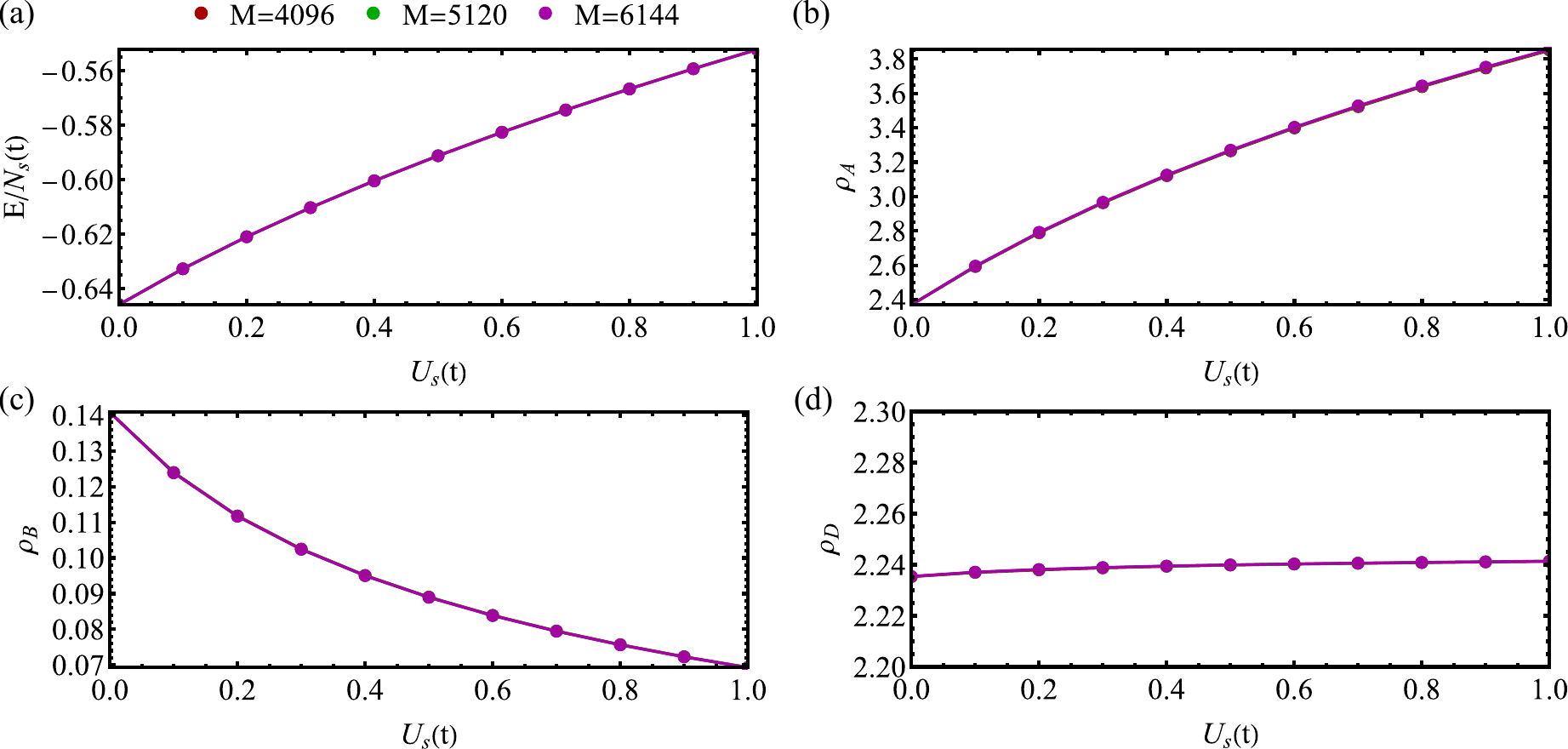}
\caption{Results of DMRG calculation at different bond dimensions: (a) ground state energy $E/N_s$, (b) extended $s$-wave pair density $\rho_A$, (c) $s$-wave pair density $\rho_B$ and (d) $d$-wave pair density $\rho_D$ versus strength $U_s$ of projection term $\hat{H}_{\text{proj}}$ under $U=6t$, $t'=0$ with the system's length $L_x=16$. Red, green, and purple colors correspond to the bond dimensions $M=4096, 5120, 6144$.}
\label{Fig:conv(U6t'0)16*4}
\end{figure}

\begin{figure}[htbp]
\centering
\includegraphics[width=0.9\linewidth]{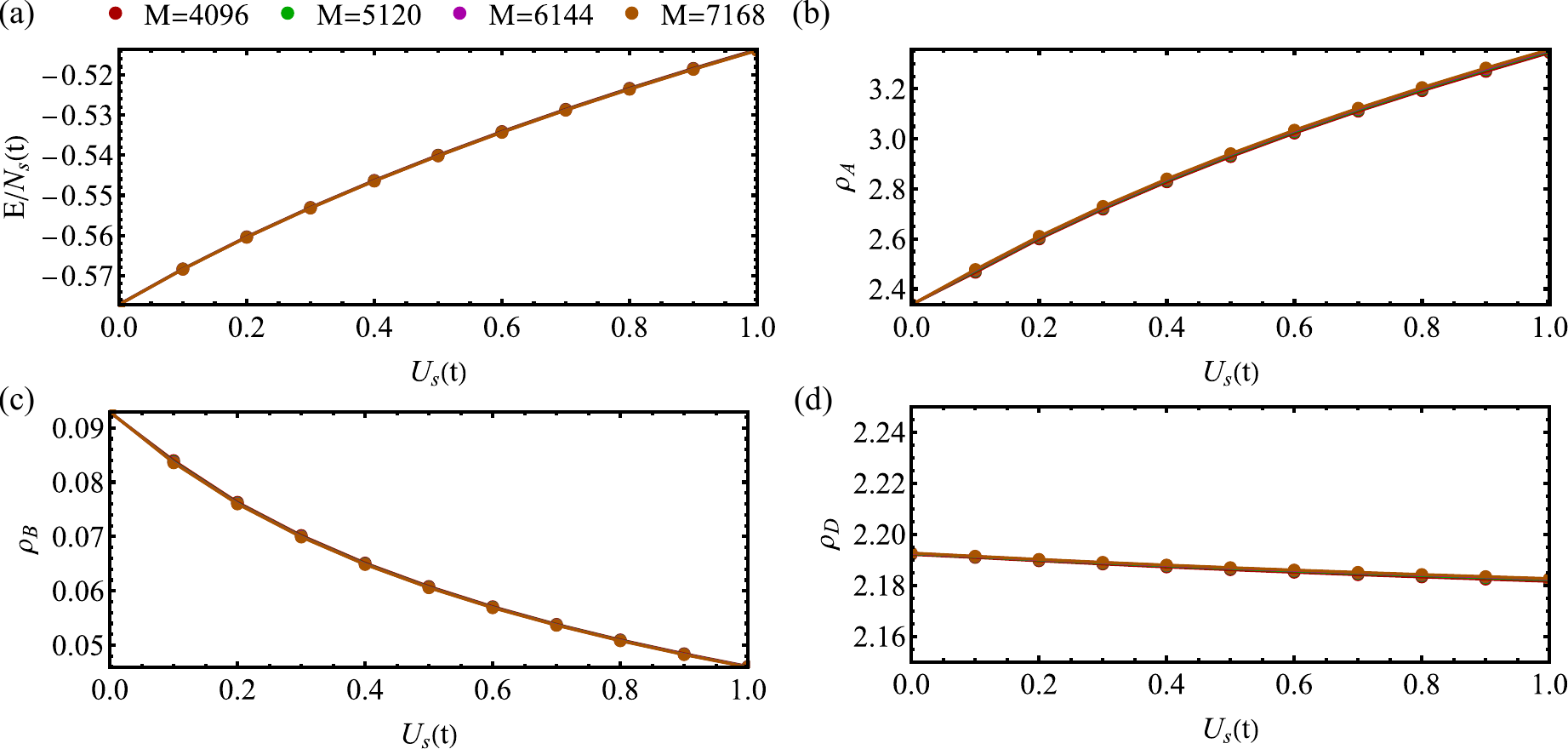}
\caption{Results of DMRG calculation at different bond dimensions: (a) ground state energy $E/N_s$, (b) extended $s$-wave pair density $\rho_A$, (c) $s$-wave pair density $\rho_B$ and (d) $d$-wave pair density $\rho_D$ versus strength $U_s$ of projection term $\hat{H}_{\text{proj}}$ under $U=8t$, $t'=0$ at hole doping $\delta=1/32$ with the system's length $L_x=16$. Red, green, purple, and brown colors correspond to the bond dimensions $M=4096, 5120, 6144, 7168$.}
\label{Fig:conv(2ht'0)16*4}
\end{figure}

\begin{figure}[htbp]
\centering
\includegraphics[width=0.9\linewidth]{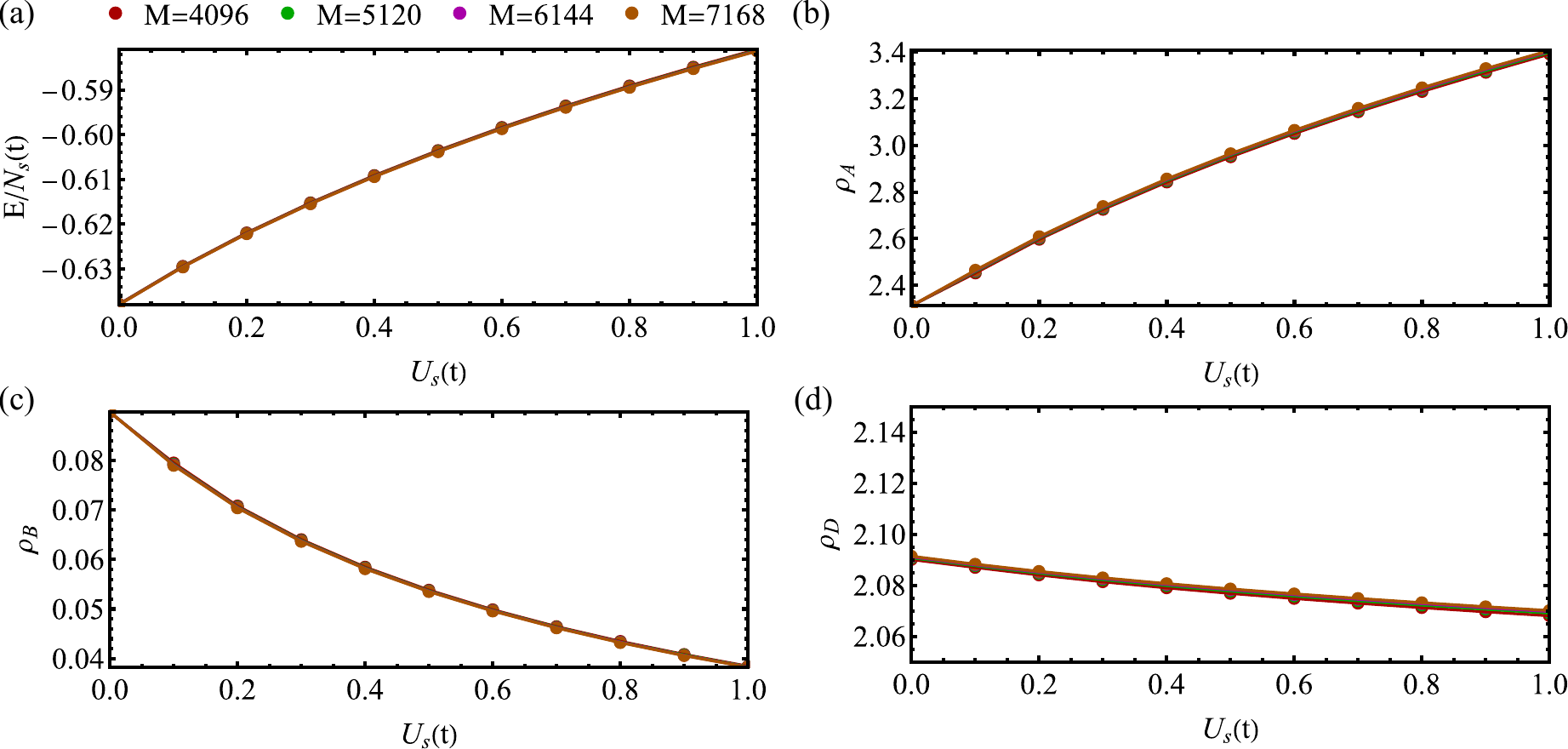}
\caption{Results of DMRG calculation at different bond dimensions: (a) ground state energy $E/N_s$, (b) extended $s$-wave pair density $\rho_A$, (c) $s$-wave pair density $\rho_B$ and (d) $d$-wave pair density $\rho_D$ versus strength $U_s$ of projection term $\hat{H}_{\text{proj}}$ under $U=8t$, $t'=0$ at hole doping $\delta=1/16$ with the system's length $L_x=16$. Red, green, purple, and brown colors correspond to the bond dimensions $M=4096, 5120, 6144, 7168$.}
\label{Fig:conv(4ht'0)16*4}
\end{figure}

\end{document}